\documentclass[11pt]{article}
\pdfoutput=1

\usepackage[utf8]{inputenc}
\usepackage{multirow}
\usepackage{amsmath, amsfonts, amssymb}
\usepackage{comment}
\usepackage{graphicx}
\usepackage{psfrag}
\usepackage{amsthm}
\usepackage[usenames,dvipsnames,svgnames,table]{xcolor}
\usepackage{enumerate}
\usepackage{arydshln}
\usepackage{soul}
 \usepackage{slashed}
\usepackage{subfig}
\usepackage{mathrsfs}
 \usepackage{a4wide}
  \usepackage{tikz}
  \usepackage{tikz-cd}
  \usetikzlibrary{shapes.geometric}
  \usepackage{tcolorbox}
\usepackage{xcolor}
\usepackage{listings}

\definecolor{codebg}{rgb}{0.97,0.97,0.97}
\definecolor{codeframe}{rgb}{0.85,0.85,0.85}
\definecolor{keyword}{rgb}{0.15,0.15,0.7}
\definecolor{string}{rgb}{0.65,0.15,0.15}
\definecolor{comment}{rgb}{0.25,0.5,0.35}
\definecolor{number}{rgb}{0.55,0.0,0.55}
\definecolor{identifier}{rgb}{0.1,0.1,0.1}

\lstdefinestyle{pythonstyle}{
    language=Python,
    backgroundcolor=\color{codebg},
    basicstyle=\ttfamily\scriptsize,
    keywordstyle=\color{keyword}\bfseries,
    stringstyle=\color{string},
    commentstyle=\color{comment}\itshape,
    numberstyle=\tiny\color{number},
    identifierstyle=\color{identifier},
    frame=single,
    rulecolor=\color{codeframe},
    showstringspaces=false,
    breaklines=true,
    breakatwhitespace=true,
    tabsize=4,
    numbers=left,
    numbersep=6pt,
    xleftmargin=1.5em,
    framexleftmargin=1.5em,
    frameround=tttt
}

\lstnewenvironment{pythoncode}[1][]{
    \lstset{style=pythonstyle,#1}
}{}

\lstnewenvironment{Mathematicacode}[1][]{
    \lstset{
    language=Mathematica,
    basicstyle=\small\ttfamily,
    keywordstyle=\color{blue},
    commentstyle=\color{gray},
    stringstyle=\color{red},
    breaklines=true,
    frame=single
}
}{}

  \usepackage{color}
  \definecolor{dark-gray}{gray}{0.20}
  \definecolor{gray}{gray}{0.30}
  \definecolor{light-gray}{gray}{0.80}
  \definecolor{dark-red}{rgb}{0.7,0,0}
  \definecolor{dark-green}{rgb}{0.1,0.4,0}
  \definecolor{dark-blue}{rgb}{0.3,0.3,0.7}
  \definecolor{light-blue}{rgb}{0.8,0.8,1}
      \definecolor{swamp}{RGB}{240, 199, 197}

  \usepackage{pifont}

\usepackage{newunicodechar} % To write the definition on the next line
\usepackage{setspace}

\newcommand{\be}{\begin{equation}}
\newcommand{\ee}{\end{equation}}
\newcommand{\eq}[1]{(\ref{#1})}
\def\be{\begin{equation}}
\def\ee{\end{equation}}
\def\bea{\begin{eqnarray}}
\def\eea{\end{eqnarray}}

\newcommand{\beq}{\begin{equation}}  \newcommand{\eeq}{\end{equation}}
\newcommand{\bal}{\begin{aligned}}   \newcommand{\eal}{\end{aligned}}
\def\beqa{\begin{eqnarray}}
\def\eeqa{\end{eqnarray}}

\newcommand{\C}{\mathbb{C}}
\newcommand{\R}{\mathbb{R}}

\newcommand{\Z}{\mathbb{Z}}
\newcommand{\Spin}{\mathrm{Spin}}

\newcommand{\Hom}{\mathrm{Hom}}

\def\simleq{\; \raise0.3ex\hbox{$<$\kern-0.75em
      \raise-1.1ex\hbox{$\sim$}}\; }
   \def\simgeq{\; \raise0.3ex\hbox{$>$\kern-0.75em
      \raise-1.1ex\hbox{$\sim$}}\; }

\numberwithin{equation}{section}

\usepackage{jheppub}
\usepackage{hyperref}
\usepackage{cleveref}

\hypersetup{
	colorlinks=true,
	linkcolor=dark-blue,
	citecolor=dark-red,
	urlcolor=dark-green,
	linktoc=page
}

\theoremstyle{remark}

\crefname{appendix}{Appendix}{Appendices}

\title{\centering Asymmetric orbifolds with vanishing\\ one-loop vacuum energy}

\author[1,2]{Vittorio Larotonda,}
\author[3]{Miguel Montero}
\author[3]{and Michelangelo Tartaglia}
\affiliation[1]{Dipartimento di Fisica e Astronomia, Universit\`{a} di Bologna, via Irnerio 46, Bologna, Italy}
\affiliation[2]{INFN, Sezione di Bologna, viale Berti Pichat 6/2, Bologna, Italy}
\affiliation[3]{Instituto de Física Teórica IFT-UAM/CSIC, C/ Nicolas Cabrera 13-15, Campus de Cantoblanco, 28049 Madrid, Spain}

\emailAdd{vittorio.larotonda@unibo.it}
\emailAdd{miguel.montero@csic.es}
\emailAdd{michelangelo.tartaglia@estudiante.uam.es}

\abstract{
We present a systematic study of non-supersymmetric type II toroidal asymmetric orbifolds with vanishing vacuum energy at one-loop in string perturbation theory. These are engineered through the conservation of a supercharge-like operator in each individual sector in the orbifold sum, despite the overall explicit breaking of spacetime SUSY. We provide a full classification of such orbifolds with finite Abelian point-group, which can only admit $\Z_k \times \Z_k$ point group with $k=2,3,4$. We present detailed constructions, alongside other examples with non-Abelian point group. For some of these models, it is possible that this cancellation persists at higher loops.}

\hypersetup{pageanchor=false}
\makeatletter
\let\old@fpheader\@fpheader
\preprint{IFT-26-11}

\hypersetup{
    pdftitle={A landscape of Non-Supersymmetric asymmetric orbifolds with vanishing Cosmological Constant},
    pdfauthor={Vittorio Larotonda, Miguel Montero, Michelangelo Tartaglia},
    pdfsubject={asymmetric orbifolds}
}

\newcommand{\ben}{\begin{enumerate}}
\newcommand{\een}{\end{enumerate}}

\begin{document}

\maketitle

\section{Introduction: Orbifolds and vanishing cosmological constant}\label{sec:orbifolds_with_vanishing_CC}

In string perturbation theory, the vacuum energy has an expansion in powers of $g_s$, just like any other observable. In a theory of closed strings, this takes the form
\begin{equation} V= \sum_{g=0}^{\infty} V_g\, g_s^{2g-2}.\label{vace}\end{equation}
The $g=0$ term is always absent in perturbative string theory due to $SL(2,\mathbb{C})$ invariance. The $g=1$ contribution is of the form
\begin{equation}\label{eq:full_partition_function}
   V_1 \sim \int_\mathcal{F} \frac{d^2\tau}{(\text{Im}{\tau})^2} \mathcal{Z}(\tau,\bar{\tau}),\end{equation}
where $\mathcal{Z}(\tau,\bar{\tau})$ is the torus partition function of the worldsheet CFT, taking into account the GSO projection, and
\begin{equation}
    \mathcal{F} = \left\{ \tau \in \C \,|\,\tau_1 \in [-\tfrac{1}{2},\tfrac{1}{2}], \tau_2>0,|\tau|>1  \right\}
\end{equation}
is the fundamental domain of $SL(2;\Z)$.

In this paper, we focus on theories where one or more terms in the expansion \eq{vace} vanish, leading to very small or zero vacuum energies in perturbation theory. We are particularly interested in mechanisms ensuring that $V_1$ vanishes, since this is the leading correction to the vacuum energy and also it is the most explored in the literature as we will now review.  

The simplest way to make the $V_g$ vanish is unbroken supersymmetry in the target space. When the target space compactification has four or more supercharges (as is the case for any supersymmetric compactification to four or more dimensions), supersymmetry is  enough to guarantee that perturbative corrections to the vacuum energy vanish, so $V_g=0$ for all $g$ due to Bose-Fermi degeneracy. In perturbative string theory, this was first argued in \cite{Martinec:1986wa}, though making the argument rigorous takes much more work \cite{Witten:2012bh,Witten:2013cia}.  The absence of corrections to the vacuum energy is most easily understood as a consequence of spacetime non-renormalization theorems. For instance, in a 4d $\mathcal{N}=1$ model, the non-renormalization theorems of the superpotential \cite{Dine:1987xk,Witten:2013cia} guarantee that no F-terms can be generated by loop corrections. Since no F-terms can appear, supersymmetry can only be broken via a D-term, like a Fayet-Iliopoulos term. This can be generated only in heterotic string compactifications to four dimensions at one-loop, and is related to the presence of a Green-Schwarz coupling. In this case, the breaking of supersymmetry is in some sense topological, and in all known examples it can be restored by shifting the vev of fields parametrizing the vacuum appropriately. At any rate, here we will focus on perturbative type II models, where the above does not occur.

In the absence of spacetime supersymmetry, the options for producing a vanishing cosmological constant are more limited, and the existing mechanisms apply to one or various terms in the expansion \eq{vace}, rather than holding at all orders.  Attaining a vanishing $V_1$ (or even at higher orders) without supersymmetry is particularly appealing since it may provide a fully stringy mechanism yielding small vacuum energies (see \cite{Angelantonj:2003hr, Dudas:2025yqm} for open-string realizations). The existing proposals in the literature are of two kinds: \begin{itemize}
\item The integrand $\mathcal{Z}(\tau,\bar{\tau})$ may be non-zero, but the full integral over the moduli space in \cref{eq:full_partition_function} still vanishes. This is the case of Atkin-Lehner symmetry \cite{Moore:1987ue}, which however has only been realized in two-dimensional target spaces so far \cite{Dienes:1990qh}. Spectra of particle theories with similar cancellations have also been engineered \cite{Satoh:2021nfu}, based on type II or heterotic toroidal asymmetric orbifold constructions \cite{Satoh:2016izo}. Atkin-Lehner symmetry cannot be realized in the context of toroidal orbifolds for target space dimension greater than two \cite{Gannon:1992su}.  
\item One could also have some symmetry reason so that $\mathcal{Z}(\tau,\bar{\tau})=0$  at each value of $\tau$ -- just as it would happen if the theory was supersymmetric. 
\end{itemize}
The focus of this paper will be on this second class of models. 

The pioneering idea to engineer models where the one-loop partition function $\mathcal{Z}(\tau,\bar{\tau})$ vanishes identically without supersymmetry comes from \cite{Kachru:1998hd}. The basic idea presented in that paper is as follows: consider a worldsheet CFT,  obtained as an orbifold of some other ``parent'' CFT by a finite group $G$. The partition function for the orbifold theory can be written as a sum 
\begin{equation}\label{eq:orbifold_partition_function}
    \mathcal{Z}_{\text{orbifold}}(\tau,\bar{\tau}) = \frac{1}{N} \sum_{(f,g)} \mathcal{Z}[f,g]\quad \text{with} \quad  f,g\in G\quad\text{and}\quad fgf^{-1}g^{-1}=1\,.
\end{equation}
where $\mathcal{Z}[f,g]$ denotes the torus partition function of the parent CFT with holonomies of $f\in G$ along the $a$-cycle of the torus and $g\in G$ along the $b$-cycle\footnote{The restriction of the sum to commuting pairs $(f,g)$ comes from the fact that orbifolding the symmetry $G$ means gauging it, and principal $G$-bundles over $T^2$ for discrete groups correspond to maps from $\pi_1(T^2)=\mathbb{Z}^2$ to the group $G$. These are described by the image of the generators, which are precisely commuting pairs of elements.}. Finally, $N$ is a normalization factor.

In each sector, the partition function $\mathcal{Z}[f,g]$ has a Hilbert space interpretation:
\begin{equation}
    \mathcal{Z}[f,g]=\text{Tr}_{\mathcal{H}_f} (q^{L_0-c_L/24} \bar{q}^{\bar{L}_0-c_R/24}\hat{g}\, P_{GSO})
\end{equation}
where the trace is taken over the $f$-twisted Hilbert space. If there is a spacetime supercharge operator in the $f$-twisted sector which commutes with $\hat{g}$, the partition $\mathcal{Z}[f,g]$ will vanish because of Bose-Fermi cancellation. 

Now imagine there is a choice of CFT and $G$ such that each commuting pair $(f,g)$ preserves some supercharge, but crucially, \emph{not the same} in all sectors (see \cite{Kachru:1998hd, GrootNibbelink:2017luf}). Then $\mathcal{Z}_{\text{orbifold}}(\tau,\bar{\tau})$ will vanish even though no supersymmetry is actually present in the theory. 

In practice, it is easiest to engineer models like this when the only commuting pairs are of the form
\begin{equation}
    (g^a,g^b) \quad \text{with } \, a,b\, \in\mathbb{Z}
\end{equation}
with $g$ running over $G$. These are all holonomies in a modular orbit of $(1,g)$ -- that is, they are related to each other by $T^2$ large diffeomorphisms. In this case, to ensure vanishing of $\mathcal{Z}_{\text{orbifold}}(\tau,\bar{\tau})$ it is enough to check whether for each $g \in G$ the action $\hat{g}$ in the untwisted Hilbert space preserves some supercharge. Then $\mathcal{Z}[1,g]$ vanishes because of Bose-Fermi cancellation, and by modular covariance the full orbit $\mathcal{Z}[g^a,g^b]$ does as well.

As mentioned above, the mechanism just described for ensuring vanishing vacuum energy at one loop was introduced in  \cite{Kachru:1998hd}. This same reference also provided a concrete asymmetric toroidal orbifold $T^6/(\mathbb{Z}_2\times\mathbb{Z}_2)$ with vanishing 1-loop vacuum energy, which however does \emph{not} quite realize the mechanism explained above. In particular, since the orbifold group is $\mathbb{Z}_2\times\mathbb{Z}_2$, it contains a non-trivial commuting pair $(f,g)$ with $f,g$ generating the two $\Z_2$ factors. In the example studied in \cite{Kachru:1998hd}, this extra term just happens to vanish, as explained in \cite{Harvey:1998rc} and reviewed below. As we will concretely see when constructing our models, this vanishing is not general at all.  The incorrect claim in \cite{Kachru:1998hd} that the  $T^6/(\mathbb{Z}_2\times\mathbb{Z}_2)$ did not include non-trivial commuting pairs was based on an attempted reformulation of the model as a quotient of a non-compact $\mathbb{R}^6$ CFT by a non-compact extension of  the asymmetric $\mathbb{Z}_2\times \mathbb{Z}_2$ by discrete translations. As explained in \cite{Aoki:2003sy}, this can be done for symmetric toroidal orbifolds, where the symmetry being orbifolded exists for any value of the radii of the parent torus, but not for asymmetric ones, where the orbifold group is only a symmetry at specific values of the volume, since it mixes winding and Kaluza-Klein momenta non-trivially. The simplest example of this phenomenon is T-duality for a circle compactification in bosonic string theory, which is only a symmetry at the self-dual radius $R=\sqrt{\alpha'}$.

Since then, other models with similar properties \cite{Kachru:1998yy, Shiu:1998he, Blumenhagen:1998uf, Angelantonj:1999gm, Angelantonj:2004cm, Satoh:2015nlc, Aoyama:2020aaw} have been constructed, but the original mechanism as described above (using finite groups in toroidal setups) remains unrealized. In this paper we will introduce a class of asymmetric type II orbifolds that realize this mechanism. We classify all possible Abelian point group that yield finite order toroidal asymmetric orbifolds with a one-loop vanishing vacuum energy. We also discuss a couple of non-Abelian examples.

These models, despite being nonsupersymmetric, are guaranteed to be tachyon free. This is a direct consequence of the vanishing of the 1-loop partition function. If there was a tachyon in the physical spectrum, the 1-loop partition function would automatically diverge\footnote{As a function of $\tau \in \C$, the partition function diverges even if there are tachyons which are not level matched. When integrating over the fundamental domain $\mathcal{F}$, the $\tau_1$ integration imposes level matching, thus getting rid of these additional contributions. If there were level-matched tachyons however, the divergence would persist also in the integral over the fundamental domain. This is what we are interested in.}, since a term
\begin{equation}
    \mathcal{Z}(\tau,\bar{\tau}) \sim e^{-2 \pi m^2 \tau_2}
\end{equation}
diverges in the $\tau_2\rightarrow \infty$ limit for $m^2<0$ and cannot be canceled by any fermionic contribution. The absence of tachyons can be checked case by case in any of our models, but it is guaranteed a priori by the $\mathcal{Z}(\tau,\bar{\tau}) = 0$ condition.\footnote{We thank Salvatore Raucci for pointing out this shortcut to us.} As a consequence, if these worldsheet CFT's admit marginal deformations so that the model is part of a larger moduli space, the scalar potential (excluding the dilaton) can only have flat or stabilized directions. These vacua therefore may provide an elegant way to stabilize almost all moduli, similarly to what happens in string islands \cite{Dabholkar:1998kv} in the SUSY context. 

The question about existence of these kind of orbifolds is not purely academic, since \cite{Kachru:1998yy} also gave arguments that models satisfying the above properties should also have vanishing vacuum energy at two loops. The original $T^6/(\mathbb{Z}_2\times\mathbb{Z}_2)$ model was found to actually have a non-vanishing two-loop cosmological constant in \cite{Aoki:2003sy}. In fact, this reference attributed the non-vanishing 2-loop result to the fact that the partition function \cite{Kachru:1998hd} cannot be written as a sum of supersymmetric contributions in the manner described above. While we do not know if these are sufficient to ensure two-loop vanishing, the models we construct here will fill in this gap. 

The rest of the paper is organized as follows: in \hyperref[sec:asymmOrb]{Section 2} we review the techniques associated with asymmetric toroidal orbifolds; in \hyperref[sec:groupth]{Section 3} we provide a group theoretic argument which allows us to show that the only allowed Abelian toroidal orbifold point groups are $\Z_2 \times \Z_2$, $\Z_3 \times \Z_3$ and $\Z_4 \times \Z_4$; in \hyperref[sec:shifts]{Section 4} we show that to actually ensure a vanishing partition function we need to carefully add shift vectors; in \hyperref[sec:models]{Section 5} we explicitly present some new models, both with Abelian and non-Abelian point groups. Finally, in \hyperref[sec:anomalies]{Section 6} we ensure the absence of global anomalies through explicit bordism calculations, while in \hyperref[sec:conclus]{Section 7} we conclude and discuss future directions.  

\section{Asymmetric orbifolds}\label{sec:asymmOrb}

We review in this Section the main concepts and techniques related to toroidal orbifold models in string theory, focusing on their asymmetric versions. These have different actions on left and right movers, and thus have no geometric interpretation, but nonetheless give rise to sensible string theories \cite{Narain:1986qm}.
We will focus on type II theories, since in \cite{GrootNibbelink:2017luf} a systematic search of heterotic orbifolds has already excluded the existence of models which realize the vacuum energy cancellation mechanism we are interested in. There has been much recent work on asymmetric orbifolds \cite{Fischer:2012qj, Satoh:2015nlc,GrootNibbelink:2017luf,Satoh:2021nfu,Gkountoumis:2023fym, Baykara:2024vss, Gkountoumis:2025btc}, most of the material we describe below can also be found from these references.

Type II string theory compactified on a $d$-dimensional torus $T^d$ is fully characterized by specifying a Narain lattice, namely a choice of even, self-dual lattice $\Gamma^{d,d}\subset \R^{d,d}$ of signature $(d,d)$ \cite{Polchinski_1998}. Points in this lattice represent the quantized internal momenta $(p_L,p_R)$, defined in terms of Kaluza-Klein momentum and winding of the internal coordinates. The inner product is defined through the indefinite norm $p_R^2-p_L^2$.

From this data one can recover the torus lattice defining $T^d = \R^d/\Lambda_d$ as the pure winding component:
\begin{equation}\label{eq:physlattice}
    \Lambda_d = \left\{2 \pi \sqrt{\frac{\alpha'}{2}} (p_L-p_R) \, \bigg{|} \, (p_L,p_R) \in \Gamma^{d,d} \right\}\,.
\end{equation}
\\The orbifold is then obtained by quotienting the theory by a symmetry $g$ of the Narain lattice $\Gamma^{d,d}$. We focus in this work on crystalline actions, which factorize as follows: 
\begin{equation}\label{eq:Gaction}
    g | p_L,p_R \rangle = e^{2 \pi i \left(-v_L \cdot p_L + v_R \cdot p_R \right)} | \theta_L \, p_L, \theta_R \, p_R \rangle
\end{equation}
where $\theta_L$ and $\theta_R$ are both lattice automorphisms of $\Gamma^{d,d}$ usually called \textit{twists}. $\theta_L$ acts trivially on purely right-mover vectors, and vice-versa. The additional phase is encoded by a $(d+d)$-dimensional vector $v = (v_L,v_R)$ called \textit{shift vector}. In toroidal orbifolds, the full group generated by twists and shifts is often called the \textit{space group}, while the group generated only by the twists is dubbed the \textit{point group}.

We will often characterize the twist by its eigenvalues. Since the $L$ and $R$ parts are lattice automorphisms, they are ($GL(d,\mathbb{R})$- conjugate to) an integral matrix in $GL(d;\Z)$. This means that the eigenvalues come in complex conjugate pairs\footnote{If $d$ is odd, there are only $(d-1)/2$ pairs and an extra 1 eigenvalue. Since we are concretely always going to work with even $d$, we simplify the notation by writing all twists as if $d$ was even.}, and we take a representative from each pair to build the \textit{twist vector} $\Vec{\theta}$. Concretely, if $\theta$ can be diagonalized on the $L$ sector as 
\begin{equation}
    \theta_L \sim
    \begin{pmatrix}
e^{2 \pi i \theta^{L}_1} & & & & \\
& e^{-2 \pi i \theta^{L}_1} & & & \\
& & \ddots & & \\
& & & e^{2 \pi i \theta^{L}_{d/2}} & \\
& & & & e^{-2 \pi i \theta^{L}_{d/2}}
\end{pmatrix}
\end{equation}
and similarly on the right movers, we represent it as a vector:
\begin{equation}
    \Vec{\theta} = (\Vec{\theta}^L,\Vec{\theta}^R), \, \text{ with } \, \Vec{\theta}^{L/R} = (\theta^{L/R}_1,...,\theta^{L/R}_{d/2})\,.
\end{equation}

If $\theta_L = \theta_R$ then from \eqref{eq:physlattice} one can see that $g$ descends to an action on the physical torus lattice, and has thus a geometric interpretation as strings propagating on the (possibly singular) quotient space $T^d/g$, realizing a so-called symmetric orbifold. Symmetric toroidal orbifolds describe geometric compactification manifolds, as they can be viewed as sigma models where the volume is always a modulus.

On the other hand, if $\theta_L \neq \theta_R$, we lose the geometric interpretation but can nonetheless study the quotiented worldsheet CFT directly and get sensible models. This asymmetry between left and right movers will be crucial to realize vacuum energy cancellation, so from now on we focus on asymmetric orbifolds.

There are consistency conditions on the choices of twist $\theta$ and shift vector $v$, first among them modular invariance. Let us consider an action generated by an element $g$ of order $N$, i.e. such that $g^N$ acts trivially. From iterating the action \eqref{eq:Gaction} one can see that an order $N$ transformation necessarily has $\theta^N = 1$ and
\begin{equation}\label{eq:shift}
    v^* \in \frac{1}{N}I \qquad \text{ where } \qquad I = \bigg{\{} \, (p_L,p_R) \, \bigg{|} \, \theta | p_L,p_R \rangle = |p_L,p_R \rangle \, \bigg{\}}\,,
\end{equation}
and $v^*$ is the projection of $v$ on the vector space underlying the invariant lattice $I$.
What we have described so far is the action of $\theta$ on bosonic coordinates of the sigma model. The action on the fermions might generically have double the order, since a $2\pi$ rotation equals multiplication by $(-1)$ for them. To ensure this is not the case and that the full action on the theory is order $N$, one needs to impose:
\begin{equation}
    N \sum_i \theta_i^L = N \sum_i \theta_i^R = 0 \quad \text{mod 2}\,.
\end{equation}

For Abelian point groups, modular invariance is equivalent to level matching \cite{Narain:1986qm,Freed:1987qk}, which is obtained by demanding that in any sector twisted by an element of order $N$
\begin{equation} \label{eq:levelMatch}
    N (E_L - E_R) \in \Z
\end{equation}
for any pair of left and right-moving states. In type II orbifolds, this translates to a condition on the twist and shift vectors as \cite{Baykara:2024vss}
\begin{equation}\label{eq:modInv}
    \sum_i \{ \theta^R_i \} - \sum_i \{ \theta^L_i \}  + (v^*_R)^2-(v^*_L)^2 \, \in \, \frac{2\Z}{N} \,.
\end{equation}

Imposing this level matching condition for every cyclic subgroup is enough to ensure modular invariance for orbifolds by Abelian groups.
For non-Abelian point groups there are additional consistency conditions \cite{Freed:1987qk}. We examine them in detail for the non-Abelian groups of our interest in \hyperref[sec:anomalies]{Section 6}.

All of these conditions, both the Abelian and non-Abelian cases, boil down to the requirement that the symmetry we are orbifolding by is free from anomalies. The general way to study this is via the anomaly theory, as we will review in \hyperref[sec:anomalies]{Section 6}.

For our constructions with vanishing one-loop vacuum energy, we are going to use twists that realize specific representations on supercharges, as explained in \hyperref[sec:groupth]{Section 3}, and then through \eqref{eq:shift} and \eqref{eq:modInv} in \hyperref[sec:shifts]{Section 4} we find appropriate shifts that ensure modular invariance. 
We also require that the twisted sectors do not contain massless gravitini invariant under the orbifold projection. If that were the case, the theory would be SUSY, making the vanishing of the partition function a simple consequence of that.

Let us examine under which condition this SUSY restoration could happen, by specializing to orbifolds of Type II theories on $T^6$. We follow the formalism and conventions of \cite{Baykara:2023plc}.

In a sector $\mathcal{Z}[g,1]$ twisted by an element $g=(\theta,v)$, masses of states are determined by their $SO(8)$ weight vector $r$, their internal momentum $p$, and the twist and shift vectors $\theta$ and $v$. Since it is convenient to work in $SO(8)$ language in light-cone quantization, we express an $SO(6)$ twist as a vector $\theta= (0,\theta_1,\theta_2,\theta_3)$, where the first component refers to the two extended not-light cone directions, on which the orbifold does not act. The Hamiltonian of such a state on the left is given as (see \cite{Baykara:2023plc})
\begin{equation}\label{eq:twistedH}
    H_{L} = N_B + \frac{(r_L + \theta_L)^2}{2} + \frac{(p_L + v_L)^2}{2} + E_0 -\frac{1}{2}, 
\end{equation}
where $N_B$ is the number of bosonic oscillators, which we will always take to be $N_B=0$, and the zero point energy is given as
\begin{equation}
    E_0 = \sum_i \frac{1}{2} |[ \theta_{L,i}]|(1 - |[ \theta_{L,i}]|).
\end{equation}
The same formula holds on right movers, by replacing $L\leftrightarrow R$.
The GSO projection imposes restrictions on the weight vectors:
\begin{equation}
\begin{aligned}
    r_L &= \begin{cases}
                (n_1, n_2, n_3, n_4), & \sum n_i = \text{odd}, \quad \text{NS-sector}, \\[10pt]
                \left(n_1 + \frac{1}{2}, n_2 + \frac{1}{2}, n_3 + \frac{1}{2}, n_4 + \frac{1}{2}\right), & \sum n_i = \text{odd}, \quad \text{R-sector},
           \end{cases}
    \\r_R &= \begin{cases}
(n_1, n_2, n_3, n_4), & \sum n_i = \text{odd}, \quad \text{NS-sector}, \\[10pt]
\left(n_1 + \frac{1}{2}, n_2 + \frac{1}{2}, n_3 + \frac{1}{2}, n_4 + \frac{1}{2}\right), & \sum n_i = \text{odd}, \quad \text{R-sector (IIB)}, \\[10pt]
\left(n_1 + \frac{1}{2}, n_2 + \frac{1}{2}, n_3 + \frac{1}{2}, n_4 + \frac{1}{2}\right), & \sum n_i = \text{even}, \quad \text{R-sector (IIA)}\,.
\end{cases}
\end{aligned} 
\end{equation}
Internal momenta live in the dual of the invariant lattice of $g$:
\begin{equation}
    p \in I^*.
\end{equation}

For a cyclic orbifold, a convenient trick to lift all masses of twisted sector is to have the orbifold acting as a pure shift on some direction of the torus. Then for a large enough radius  there are no twisted massless states (see Appendix A of \cite{Baykara:2023plc}).
We will argue in \hyperref[sec:shifts]{Section 4}, no such orbifold with Abelian point group $P_G$ can have vanishing partition function other than the classic $\Z_2 \times \Z_2$ examples. This forces us to consider models where no radius is large, and we have to carefully analyze the twisted sectors individually to make sure that no gravitini appear. Additionally, this means that Abelian toroidal orbifolds with vanishing 1-loop vacuum energy can only exist in four dimensions or lower (since the whole $T^6$ needs to be small) and there cannot be any decompactification limit.
For a non-Abelian group, some of the radii of the parent torus may remain large, but we need to analyze other twisted sectors in detail anyway. We do not know whether non-Abelian orbifolds with a vanishing 1-loop vacuum energy exist in more than four dimensions.

%For a cyclic orbifold, a convenient trick to lift all masses of twisted sector is to have the orbifold acting as a pure shift on some direction of the torus. Then for a large enough radius it can be explicitly shown (see Appendix A of \cite{Baykara:2023plc}) that there are no twisted massless states.
%We will argue in \hyperref[sec:shifts]{Section 4}, no such orbifold with Abelian point group $P_G$ can have vanishing partition function other than the classic $\Z_2 \times \Z_2$ examples. This also cannot work for a non-Abelian group, the commutator subgroup $[G,G]$ is non-trivial, but will act trivially on any component in which the generators are purely shifts. We then have to carefully analize the twisted sectors individually, to make sure no gravitini are restored.

Fortunately, there are strict restrictions on the orbifold group elements $g$ that allow for massless twisted gravitini. This is because a 4D gravitino can only arise as a tensor product of an $SO(2)$ vector, say on the left, and an $SO(2)$ spinor, say on the right.
The highest weight of a vector representation is $r_L=(1,0,0,0)$, so we can plug it in \eqref{eq:twistedH} to give:
\begin{equation}\label{eq:Hvec}
    H_L^\text{vect} = \frac{(p_L + v_L)^2}{2} + \sum_i |\theta_{L,i}|.
\end{equation}
This can be zero only if $v_L \in I^*$, such that there is a $p$ that can cancel the first term in \eqref{eq:Hvec}, and furthermore the twist is trivial, $\theta_i = 0$. So if the orbifold group does not contain elements for which the twist is trivial on either side, we are safe from accidental SUSY restoration.

On the other side, weights of spinors are $r_R = \frac{1}{2}(\epsilon_1,\epsilon_2,\epsilon_3,\epsilon_4)$ for $\epsilon_i = \pm 1$, for which the mass is
\begin{equation}\label{eq:Hspinor}
    H_R^\text{spinor} = \frac{(p_R + v_R)^2}{2} + \frac{1}{2}(1 + \epsilon_i)\theta_i.
\end{equation}
This only puts restrictions on the shift vector $v$, since for any twist one can always find a choice of sign that kills the second term.
For our models, we will carefully choose shifts such that if there is any element for which either $\theta_L=0$ or $\theta_R=0$, the corresponding shifts lifts the ground state mass to be positive.

Lastly, we are going to use special points in the Narain moduli space where the torus lattice has an enhanced discrete symmetry. One way to build them is as follows \cite{Dabholkar:1998kv}: pick a Lie algebra $\mathfrak{g}$ of dimension $d$, and construct
\begin{equation}\label{eq:extraSymm}
    \Gamma^{d,d}(\mathfrak{g}) = \left\{ (p_L,p_R) \in \Lambda_W(\mathfrak{g}) \times \Lambda_W(\mathfrak{g}) \, \bigg{|} \, p_L-p_R \in \Lambda_R(\mathfrak{g})   \right\} \,,
\end{equation}
where $\Lambda_R(\mathfrak{g})$ and $\Lambda_W(\mathfrak{g})$ are respectively the root and weight lattice of the Lie algebra $\mathfrak{g}$. This is an even, self-dual lattice of signature $(d,d)$, and thus defines a special point in the Narain moduli space. By construction the Weyl group of $\mathfrak{g}$ acts on $\Gamma^{d,d}(\mathfrak{g})$, and therefore  is a symmetry at this point.

\section{Constraints on Abelian point groups} \label{sec:groupth}

In this Section we  approach the search of orbifolds with the properties discussed in the \hyperref[sec:orbifolds_with_vanishing_CC]{Introduction} from a purely group theoretic point of view, using the tools introduced in \hyperref[sec:asymmOrb]{Section 2}. In particular we prove that, in the context of toroidal orbifolds, the only Abelian point groups that could yield a non-supersymmetric 4D model with vanishing one-loop cosmological constant are $\mathbb{Z}_3 \times \mathbb{Z}_3$ and $\mathbb{Z}_4 \times \mathbb{Z}_4$, together with the original $\mathbb{Z}_2 \times \mathbb{Z}_2$ constructions à la \cite{Kachru:1998hd,Harvey:1998rc,Shiu:1998he}. We will then see in \hyperref[sec:shifts]{Section 4} that in the absence of shifts, these purely Abelian constructions have non-zero $\mathcal{Z}(\tau,\bar{\tau})$ in some twisted sectors. One has to be careful with introducing shifts to get rid of these sectors in the sum \eqref{eq:orbifold_partition_function}, because generically will introduce global anomalies, violating modular invariance. Before worrying about shifts, in this Section we restrict the point groups we are allowed to consider in the first place.

We will first start describing the general strategy we will follow, and then make it more concrete in the context of toroidal orbifolds. As explained in the introduction, we would want to construct orbifolds where any commuting pair of holonomies on $T^2$ preserves some supercharge.
In practice, a way to engineer this is to have the only commuting pairs to be of the form
\begin{equation}
    (g^a,g^b) \quad \text{with } \, a,b\, \in\mathbb{Z},
\end{equation}
with $g$ running over $G$. These are all holonomies in a modular orbit of $(1,g)$ -- that is, they are related to each other by $T^2$ large diffeomorphisms. In this case, to ensure vanishing of $\mathcal{Z}_{\text{orbifold}}(\tau,\bar{\tau})$, it is enough to check whether for each $g \in G$ the action $\hat{g}$ in the untwisted Hilbert space preserves some supercharge. Then $\mathcal{Z}[1,g]$ vanishes because of Bose-Fermi cancellation, and by modular covariance the full orbit $\mathcal{Z}[g^a,g^b]$ does as well. We will focus on this case -- or rather, a small generalization of it, where there are additional commuting pairs that however act trivially in some factors of our CFT (see \hyperref[app:D]{Appendix D} for details) -- in what follows, though more general possibilities have been explored in the literature \cite{Satoh:2015nlc,Aoyama:2020aaw}.

We will now embark on a search of CFT's and orbifold groups $G$ that allow us to realize the above. At a group theoretical level, we can do this by finding an orbifold group, generated by elements $g_i$,  with an action on the spacetime supercharges such that:
\begin{enumerate}
    \item Each $g_i$ has an invariant subspace, ensuring that at least one supercharge is preserved in each orbifold twisted sector.
    \item The invariant subspaces interstect only trivially, so that no supersymmetry survives in the full theory.
    \item The only commuting pairs of elements in the full orbifold group are of the form $(1,g_i)$ and its modular orbits.
    \end{enumerate}
 In practice, we will focus on toroidal orbifolds of finite order, though there are a few studies of the above ideas in e.g. Gepner models \cite{Aoyama:2020aaw}. For toroidal orbifolds, reference \cite{GrootNibbelink:2017luf} proved a no-go theorem showing that no orbifolds with the three properties above exist for heterotic compactifications. As we will see, things are quite different for type II. The intuitive reason for this is that type II has twice as many supercharges, which yields more possibilities to preserve some of them in each twisted sector. For the same reason, to have a chance at realizing this mechanism we must use asymmetric orbifold constructions: symmetric type II constructions behave, from this point of view, exactly as their heterotic counterparts.
Our results are a partial analogue of the no-go of \cite{GrootNibbelink:2017luf} in the case of Abelian point groups. We leave a full classification of non-Abelian versions of this mechanism for future work. As we show in \hyperref[sec:models]{Section 5}, one can find some examples of non-Abelian groups satisfying the properties we spell out below, but a full classification needs more powerful tools than the ones developed in this Section.
 
 For a type II compactification to four dimensions, the partition function factorizes into left and right movers, and each of the two factors has the form (see e.g. \cite{Ibanez:2012zz,Gkountoumis:2023fym}):
\begin{equation}\label{eq:partition_function_factorization}
    \mathcal{Z}[f,g]= \mathcal{Z}_{\mathbb{R}^{1,3}} \cdot \mathcal{Z}^B_{T^6}[f,g] \cdot \mathcal{Z}^F[f,g]
\end{equation}
where $\mathcal{Z}_{\mathbb{R}^{1,3}}$ is the universal factor associated to the external bosons, $\mathcal{Z}^B_{T^6}[f,g]$ is the partition function of the compact bosons, and
\begin{equation}
    \mathcal{Z}^F[f,g] = \mathcal{Z}^{NS}[f,g] - \mathcal{Z}^R[f,g]
\end{equation}
is the partition function of the worldsheet fermions.
This last factor is the only one that can vanish, since the rest are partition functions of unitary bosonic CFT's, and thus are strictly positive on their own.

Recall from the discussion above that we are looking for a group action $G$ on the space of supercharges that breaks supersymmetry, but for which in each sector of the orbifold there is a (not globally defined) supercharge-like operator, which implements Bose-Fermi cancellation and ensures the vanishing of the one-loop partition function. In practice, we will study the transformation properties of untwisted gravitino vertex operators, since spacetime supercharges are built out of these. 

Following the strategy of \cite{GrootNibbelink:2017luf}, we will start by constraining the point group $P_G$, and look for an embedding of $P_G$ into the group that acts naturally on the supercharges of the parent torus theory. In type II string theory, the supercharges can be constructed from the R-NS and the NS-R sectors independently. In light-cone gauge, each sector yields 8 real supercharges, for a total of 16 (only 16 of the 32 supercharges of type II on a torus are manifest). These supercharges transform in the $(\mathbf{8},\mathbf{1})\oplus (\mathbf{1},\mathbf{8})$ representation of $Spin(8)_L\times Spin(8)_R$. A $T^6$ compactification breaks $Spin(8)$ to $\Gamma \times Spin(2)$, where $Spin(2)$ is the massless little group in four dimensions and $\Gamma$ embeds in $SU(4)\simeq Spin(6)$. Under the decomposition $Spin(8)\rightarrow SU(4) \times Spin(2)$, we have
\begin{equation}
    \mathbf{8}\rightarrow \mathbf{4}_{+1/2},
\end{equation}
so the supercharges transform on the complex $(\mathbf{4},\mathbf{1}) \oplus (\mathbf{1},\mathbf{4})$ of $SU(4)_L\times SU(4)_R$. We are looking for a faithful\footnote{If it is not faithful, we take $P_G$ as the group that is faithfully represented by $\rho$.} representation $\rho:G\rightarrow GL(V)$ of a finite group $G$ with the following properties:
\begin{itemize}
    \item[] \textbf{I.} The representation space $V$ is complex 8-dimensional, as is the space of supercharges, and $\rho$ splits as two 4D subrepresentations, the left and right movers.
    \item[] \textbf{II.} In each 4D block, every element has determinant 1, to land in $SU(4)_L\times SU(4)_R$.
    \item[] \textbf{III.} Every element fixes some vector, i.e. $\forall g \in G, \, \exists \, v \in V$ such that $\rho(g) \cdot v = v \,$.
    \item[] \textbf{IV.} $\rho$ does not have a trivial subrepresentation, i.e. no vector is fixed by the whole $G$.
    \item[] \textbf{V.} $\rho$ corresponds to the spin lift of a crystallographic symmetry, that acts on the underlying boson lattice. In other words we demand that the action $\rho$ can be extended to bosons consistently.
\end{itemize}
To constrain this problem, we will start with representations of generators of $G$, and try to put them together without spoiling any of the above properties. Every element of $G$ generates some finite $\Z_k$ subgroup, which can be diagonalized over the complex numbers and is thus represented as a diagonal matrix with entries only $k$-th roots of unity. We represent any such generator $f$ as a vector

\begin{equation}\label{eq:f}
    f = (z^{f_1},z^{f_2},z^{f_3},z^{f_4}|z^{f_5},z^{f_6},z^{f_7},z^{f_8}), \quad \text{with}\, z = e^\frac{2 \pi i}{k},
\end{equation}
and where the vertical bar divides between the action on left and right movers. In each block, condition \textbf{II} above implies $\sum_i f_i^{L/R} = 0$ mod $k$. Furthermore, since every element of $G$ needs to fix a vector by condition \textbf{III}, at least one of the $f_i$'s needs to be 0. Without loss of generality, we will take it to be $f_1=0$ in practice. Notice that with just one generator (that is, if $G$ is cyclic), this is incompatible with \textbf{IV}, since all elements will fix the same invariant subspace as the generator. Therefore, we then look for another generator,
\begin{equation}\label{eq:g}
    g = (z^{g_1},z^{g_2},z^{g_3},z^{g_4}|z^{g_5},z^{g_6},z^{g_7},z^{g_8})
\end{equation}
and demand that the group generated by the two satisfies the properties above. With only two generators, we will already find very stringent constraints. Note that to write the second generator also in diagonal form we already assumed that it commutes with $f$ as to be simultaneously diagonalizable. Furthermore, we take it to be of the same order of $f$. This is actually generic, since if the orders of the two generators are coprime, by Sunzi's Remainder Theorem $\Z_{k_1} \times \Z_{k_2} \simeq \Z_{k_1 k_2}$ so the group is actually cyclic, and thus if an element preserves a vector then the whole group does, violating condition \textbf{IV}. This means we can consider every prime factor individually, and then if the order $k_1$ of one generator divides the order $k_2$ of the other, we view them both as order $k_2$ and all considerations of this Section apply.

Again, in this Section we limit our considerations to Abelian point groups, since the representation theoretic tools we develop here are strong enough to fully classify the candidate groups and representations.
As shown explicitly in \hyperref[sec:models]{Section 5}, there also exist non-Abelian groups with representations satisfying properties \textbf{I}-\textbf{V} above. In particular, there are groups of arbitrary high order satisfying \textbf{I}-\textbf{IV}, showing that the tools in this Section are not sufficient to fully classify non-Abelian representations.
\\In the Abelian case, the conditions above are implemented concretely by demanding that
\begin{itemize}
    \item[] \textbf{1.} $g$ has at least an entry equal to $1$, but not in a coordinate where $f$ has a 1.
    \item[] \textbf{2.} All products $f^m \, g^n$ have an entry equal to 1.
\end{itemize}
There are constraints on the order of generators from condition \textbf{2.}, since a priori different $m,n = 0,...,k-$1 impose different conditions on the exponents $f_i,g_i$. Intuitively, the higher the order of $f,g$, the more constraints the representation needs to satisfy. The precise counting goes as follows. For every pair $(m,n)$ we are looking for a pair of exponents $(f_i,g_i)$ such that
\begin{equation} \label{eq:linearcond}
    m f_i + n g_i = 0 \quad \text{mod }k \,.
\end{equation}
This condition is invariant under rescaling of either $(m,n)$ or $(f_i,g_i)$ by a constant $\lambda \in \Z_k$, so it is convenient to view it geometrically: for every $(m,n)$ in the projective space $\mathbb{P}^1(\Z_k)$, i.e. the set of lines in $\Z_k^2$, we need a vector $v_i= (f_i,g_i) $ whose inner product with $(m,n)$ vanishes. Furthermore, a given pair $(f_i,g_i)$ cannot satisfy the conditions for more than one line simultaneously, since then $f_i=g_i=0$ which violates condition \textbf{IV}. There is thus an independent condition for each element in $\mathbb{P}^1(\Z_k)$. The dimension of this space is known: by using a prime decomposition
\begin{equation}
    k = \prod_i p_i^{\alpha_i}\,,
\end{equation}
the size of the corresponding projective line is\footnote{\label{footnote2}The counting goes as follows: from the Sunzi's Remainder Theorem we can split $\Z_k$ into its coprime factors $\Z_{p_i^{\alpha_i}}$, so we only need to compute this for $p^\alpha$. In $\left(\Z_{p^\alpha}\right)^2$, proper lines are generated by  primitive vectors (i.e. those whose components are not both divisible by p). Their number is computed as the difference between the total number of vectors, $p^{2\alpha}$, and the not primitive ones, which are all written as $p$ times the vectors in $\left(\Z_{p^{\alpha-1}}\right)^2$, so there are $p^{2(\alpha-1)}$ of them. To find the number of independent lines we need to quotient this by units in $\Z_{p^\alpha}$, whose number is the Euler totient function $\phi(p^\alpha) = p^\alpha-p^{\alpha-1}$. Putting it all together: $$\left| \mathbb{P}^1(\Z_{p^\alpha}) \right| = \frac{p^{2\alpha}-p^{2(\alpha-1)}}{\phi(p^\alpha)} = p^\alpha + p^{\alpha-1}$$}
\begin{equation} \label{eq:projective}
     \left| \mathbb{P}^1(\Z_k) \right| = \prod_{p_i^{\alpha_i}\big{|}k} (p_i^{\alpha_i}+p_i^{\alpha_i-1})\,.
\end{equation}
For $k$ prime, this simplifies to 
\begin{equation} \label{eq:primeProj}
    \left| \mathbb{P}^1(\Z_k) \right| = k + 1 \quad \text{for $k$ prime}\,.
\end{equation}
Since the representation on the supercharges must be 8-dimensional, we only have 8 exponent pairs $(f_i,g_i)$ to work with to satisfy all these conditions.  Therefore $k+1 \leq 8$ immediately imposes that the maximum order of a $\Z_k \times \Z_k$ action for a prime is $k=7$. 

Shortly we are going to argue that in a toroidal orbifolds we cannot have order 5 or order 7 generators, since they will not descend to a lattice symmetry on bosons and thus violate condition \textbf{V}.
Note that this does not exclude that a more general CFT could have $\Z_5 \times \Z_5$ or $\Z_7 \times \Z_7$ orbifolds with properties \textbf{I}-\textbf{IV}.

For us, in practice, this counting means that for toroidal orbifolds all we can play with are generators of order a power of 2 or 3, or combinations thereof.
Then we can use the general formula \eqref{eq:projective} to see that
\begin{equation}
    \left| \mathbb{P}^1(\Z_k) \right| = (3 \cdot 2^{a-1}) \cdot (4 \cdot 3^{b-1}) \quad \text{for $k = 2^a \cdot 3^b$} \,, 
\end{equation}
which is lower than 8 only for $k=2,3,4$, since $k=6,8,9$ already exceed the bound.

We could also try to add a third generator $h$. The conditions we would have to satisfy are similar: in particular the equivalent of \eqref{eq:linearcond} would read
\begin{equation}
    m f_i + n g_i + l h_i = 0 \quad \text{mod }k \,.
\end{equation}
The same exact reasoning as above leads to the number of conditions being the same as the size of the projective plane $\mathbb{P}^2(\Z_k)$. By an analogous counting, for $k$ prime this is
\begin{equation}
    \left| \mathbb{P}^2(k) \right| = k^2 + k + 1 \quad \text{for $k$ prime}\,.
\end{equation}
Which is lower than 8 only for $k=2$. This means that there could be $\Z_2 \times \Z_2 \times \Z_2$ models satisfying our conditions. These are further quotients of the original $\Z_2 \times \Z_2$ models in \cite{Kachru:1998hd, Shiu:1998he}. We have checked that in all such cases, one of the three is always acting purely as $(-1)^{F_L}$, possibly with some shift. Since orbifolding by this symmetry in ten dimensions takes one from IIA to IIB, these models are related to IIA vs. IIB versions of the model in \cite{Kachru:1998hd, Shiu:1998he}; if the shift of the $(-1)^{F_L}$ action is trivial, the models are exactly equivalent to that in \cite{Kachru:1998hd, Shiu:1998he}.

With more generators, there are even more constraints. The cardinality of $\mathbb{P}^n(\mathbb{Z}_k)$ is 
\begin{equation}
    |\mathbb{P}^n(\mathbb{Z}_k)| = k^n \prod_{p|k} \left( 1 + \frac{1}{p} + \frac{1}{p^2} + \dots + \frac{1}{p^n} \right)
\end{equation}
where $p$'s are the prime factors that divide $k$. There are no non-trivial solutions for more than three generators, which shows that our classification is a complete scan of this kind of orbifolds.

Once a suitable point group $P_G$ has been determined satisfying conditions \textbf{I}-\textbf{V}, we attempt to extend it to the full group $G$ by an appropriate choice of shift vectors. For the mechanism described in \hyperref[sec:orbifolds_with_vanishing_CC]{Section 1} to work, we in fact must ensure one additional condition: 
\begin{itemize}
    \item[] \textbf{VI.} The only commuting pairs in $G$ acting non-trivially on fermions are of the form $(g^a, g^b)$ for a generic element $g$ of $G$. \end{itemize}
This is not an arbitrary simplification: we will compute explicitly the sectors $\mathcal{Z}[f,g]$ and show that they do not vanish except for $\Z_2 \times \Z_2$ models. This accidental cancellation is in fact the reason why the original model in \cite{Kachru:1998hd} worked!
Before moving on to shifts, we can find further restrictions on the point groups, by restricting our attention to toroidal orbifolds.

\subsection{Toroidal actions}

In this work we are ultimately interested in toroidal orbifolds, whose symmetries are far more restricted than a general CFT. This puts an additional constraint (condition \textbf{V}) on the possible groups. In particular, there are a priori fermionic $\Z_k \times \Z_k$ actions compatible with our conditions \textbf{I} to \textbf{IV} with $k=5,7$. We will now show that they do not descend to a bosonic action which is a symmetry of a lattice, and thus cannot be realized as a toroidal orbifold. Note that this of course does not exclude that some non-toroidal construction, along the lines of e.g. \cite{Aoyama:2021kqa}, could do the job. 

The idea goes as follows. Starting from a given $SU(4) \simeq$ Spin(6) action of a generator on the fermions, we derive the bosonic action on the torus coordinates via the covering map Spin$(6) \rightarrow SO(6)$, which in a convenient basis is given by the external product (see \cite{GrootNibbelink:2017luf}):
\begin{equation}\label{eq:spinproj}
    \begin{aligned}
        \Lambda^2:SU(4) &\longrightarrow SO(6) \\
        f &\longmapsto f \wedge f\,.
    \end{aligned}
\end{equation}
For diagonal $f$ this takes a particularly simple form:
\begin{equation}
    \begin{aligned}
        (e^{ 2 \pi i\phi_1},e^{ 2 \pi i\phi_2},e^{ 2 \pi i\phi_3},e^{ 2 \pi i\phi_4})& \\ \longmapsto (e^{ 2 \pi i(\phi_1 + \phi_2)},e^{ 2 \pi i (\phi_3 + \phi_4)} &,e^{ 2 \pi i(\phi_1 + \phi_3)},e^{ 2 \pi i(\phi_2 + \phi_4)},e^{ 2 \pi i(\phi_1 + \phi_4)},e^{ 2 \pi i(\phi_2 + \phi_3)})\,.
    \end{aligned}
\end{equation}
This mapping can be seen to be conjugate to a real matrix, since det$f=1$ implies $\sum_i \phi_i = 0$.
\\The three independent eigenvalues of the resulting $SO(6)$ action are then read off as:
\begin{equation}\label{eq:spinpush}
    \theta_1 = \phi_1 + \phi_2, \quad \theta_2 = \phi_1+\phi_3, \quad \theta_3 = \phi_1 + \phi_4 \,.
\end{equation}
The inverse procedure is also easy to describe for diagonal matrices: one has to essentially invert \eqref{eq:spinpush}. One solution is to take (see e.g. \cite{Sen:1995ff,Baykara:2024vss} for the $SO(8)$ version of this in a stringy context):
\begin{equation} \label{eq:spinlift}
    (\phi_1,\phi_2,\phi_3,\phi_4) = \frac{1}{2}\left(
    \begin{array}{ccc}
         1 & 1 & 1 \\
         1 & -1 & -1 \\
         -1 & 1 & -1 \\
         -1 & -1 & 1 \\
    \end{array}
    \right)
    \left(
    \begin{array}{c}
         \theta_1 \\
         \theta_2 \\
         \theta_3 \\
     \end{array}
    \right) \, .
\end{equation}

The possible values of $\theta_i$ are very constrained for crystallographic symmetries, as is explained in \hyperref[app:A]{Appendix A}.
In particular, for a $\Z_5$ or a $\Z_7$ symmetry there is essentially only one option for each. Let us examine them.

For an order 5 generator, we see from \eqref{eq:cyclotomic} that the eigenvalues must be all the non-trivial 5th roots of unity, so a $\Z_5$ symmetry acting on a rank 6 lattice must necessarily act trivially on 2 directions and with eigenvalues 
\begin{equation}
    (\omega,\omega^2,\omega^3,\omega^4) \qquad \text{for} \qquad \omega = e^{2 \pi i/5}
\end{equation}
in the remaining 4. In terms of the twists $\theta_i$ this naively means
\begin{equation}
    (\theta_1,\theta_2,\theta_3) = \left(\frac{1}{5}, \frac{2}{5}, 0 \right)
\end{equation}
This lifts through \eqref{eq:spinlift} to an order 10 action on fermions, but twisting with an extra\footnote{There is a double cover ambiguity when lifting a bosonic action to fermions. The choice of lift through Spin(6)$\rightarrow$SO(6) can sometimes result in a doubling of the order of the fermionic action. This is resolved by taking twists valued in $0\leq \theta_i < 2$ instead of the naive $0\leq \theta_i < 1$. In the following we will always implicitly make these choices to have an honest $\Z_k$ action, and to satisfy modular invariance \eqref{eq:modInv}, see e.g. \cite{Shiu:1998he,Kachru:1998yy,Harvey:1998rc}} $(-1)^F$ produces an honest order 5 fermionic action as
\begin{equation} \label{eq:liftz5}
    (\phi_1,\phi_2,\phi_3,\phi_4) = \frac{1}{2}\left(
    \begin{array}{ccc}
         1 & 1 & 1 \\
         1 & -1 & -1 \\
         -1 & 1 & -1 \\
         -1 & -1 & 1 \\
    \end{array}
    \right)
    \left(
    \begin{array}{c}
         6/5 \\
         2/5 \\
         0 \\
     \end{array}
    \right)
    = \left( \frac{4}{5},\frac{2}{5},\frac{3}{5},\frac{1}{5} \right)\,.
\end{equation}

This means that to build generators $f,g$ of a toroidal $\Z_5 \times \Z_5$ action we can use as 4D left/right moving blocks either a trivial action, or one with the exponents in \eqref{eq:liftz5}. Since \eqref{eq:liftz5} does not act trivially on any supercharge, one can test the few available combinations and see that such actions either have trivial subrepresentations or elements that do not fix any vectors.

For $\Z_7$ actions the reasoning is similar: any non-trivial order 7 element needs to act on a 6D lattice with eigenvalues
\begin{equation}
    (\xi,\xi^2,\xi^3,\xi^4,\xi^5,\xi^6) \qquad \text{for} \qquad \xi = e^{2 \pi i/7} \,.
\end{equation}
We take the appropriate spinorial lift and obtain the phases:
\begin{equation}\label{eq:liftz7}
    (\phi_1,\phi_2,\phi_3,\phi_4) = \left( \frac{3}{7},\frac{5}{7},\frac{6}{7},0\right) \,, 
\end{equation}
so again we can build generators using either \eqref{eq:liftz7} or a fully trivial action. Since \eqref{eq:liftz7} has a trivial phase it looks like a more promising candidate, but again a short scan of the possible combinations shows that it is impossible to obtain local but not global SUSY preservation.

\section{Adding shifts}\label{sec:shifts}

We showed in \hyperref[sec:groupth]{Section 3} the only potential Abelian point groups that realize local SUSY are $G = \Z_2\times \Z_2, \Z_3\times \Z_3$, $\Z_4 \times \Z_4$, or $ \Z_2\times \Z_2 \times \Z_2$ . To construct a full orbifold action, these have now to be decorated by shifts, as explained in \hyperref[sec:asymmOrb]{Section 2}.

The simplest choice one could make is to have the zero vector as a shift, which we call ``no shift''. As we mentioned at the beginning of \hyperref[sec:groupth]{Section 3}, however, $\Z_3\times \Z_3$ and $\Z_4 \times \Z_4$ models without shifts do not have a vanishing vacuum energy. In this case, all group elements commute, and so the one-loop vacuum energy receives contributions from all ``mixed sectors'' like $\mathcal{Z}[f,g]$, where $f,g$ are the generators of the two cyclic factors.  There is no supercharge preserved by $f,g$ together, so we would expect that these contributions are non-vanishing. We have checked this is the case in all explicit examples. 

These worrisome mixed sectors appear because by construction the two generators commute. A natural solution to get rid of these would be to add shift vectors in such a way that the actual orbifold group becomes non-Abelian. If the generators $f,g$ are rendered not commuting, the corresponding troublesome mixed sectors above will not enter in \eqref{eq:orbifold_partition_function}. This is the same idea suggested in \cite{Kachru:1998hd}. Importantly, we wish to work with finite groups, where the orbifolding procedure is well-defined. Shifts do not affect the fermionic factor in any given sector of the partition function (see the discussion around equation \eqref{eq:partition_function_factorization}), so any sector that vanished due to the action of supercharges on fermions will still vanish after adding shifts. 

As we will see in  this Section, generic shifts will lead to models that are either not modular invariant, or that have gravitini coming from the twisted sectors.
Nevertheless, some point groups admit a consistent choice of shifts realizing the local SUSY mechanism described in the \hyperref[sec:orbifolds_with_vanishing_CC]{Introduction}.

\subsection{$\mathbb{Z}_2\times \mathbb{Z}_2$ point group}
We start by considering the $\Z_2 \times \Z_2$ constructions of \cite{Kachru:1998hd}. 
In our language, the original models have twist vectors
\begin{equation}
    \begin{aligned}
        f:& \quad &\theta_f^L &= \left( \frac{1}{2},\frac{1}{2},0\right), \, &\theta_f^R &= (0,0,0) \,, \\
        g:& \quad &\theta_g^L &= (0,0,0), \, &\theta_g^R &= \left( \frac{1}{2},\frac{1}{2},0\right) \,.
    \end{aligned}   
\end{equation}
Using formulas from \hyperref[app:TorusZ]{Appendix B} we can compute explicitly on the right:
\begin{equation}
    \begin{aligned}
        \mathcal{Z}_{NS}[f,g] = \frac{1}{2\eta^4}\left(\vartheta_3^2\vartheta_4^2-\vartheta_4^2\vartheta_3^2\right) = 0\,, \\
        \mathcal{Z}_{R}[f,g] = \frac{1}{2\eta^4}\left(\vartheta_2^2\vartheta_1^2+\vartheta_1^2\vartheta_2^2\right) = 0\,.
    \end{aligned}
\end{equation}
As a result, the one-loop vacuum energy vanishes in this model, including the dangerous ``mixed sectors'' that we discussed. However, this cancellation is a peculiar feature of the $\mathbb{Z}_2\times \mathbb{Z}_2$ model, and in fact is not generic. At an algebraic level, it works because the $\vartheta$-functions transform into one another, but this works only twists are order 2. This means that the original models \cite{Kachru:1998hd,Kachru:1998yy, Shiu:1998he} do not quite realize the local SUSY structure described in the previous Section, but rather rely on an ``accidental'' cancellation which will not happen in more general models, as we will now see.

\subsection{$\Z_3 \times \Z_3$ point group}

For an Abelian point group, all generators can be diagonalized simultaneously. For order 3 rotations, this means that the action factorizes nicely on the torus as $T^6 \simeq T^2 \times T^2 \times T^2$. For a cyclic orbifold, the easiest way to get rid of massless twisted sectors is to consider a point group action which is trivial on the last $T^2$. Then one can make the orbifold freely acting by adding a shift on these directions, and by taking the torus to be large enough this ensures that all twisted sectors are massed up (see e.g. Appendix A of \cite{Baykara:2023plc} for a detailed derivation of this fact). We will see now however that this is actually too strong a restriction on our point groups, that leads to no solutions. As a result, we will be forced to look at models where all radii have stringy size in \hyperref[sec:models]{Section 5}.

As discussed above, we must find shifts that can be added to the point-group generators to make them not commute with each other. Because of the aforementioned factorization, we can look at each 2D block individually.
For a generic rotation $R$, and a shift $S_v$ by a vector $v$, one can see that the action of their commutator is
\begin{equation}
    R S_v R^{-1} S_v^{-1} = S_{Rv-v}\,.
\end{equation}
Because of this, on a torus $T^d=\R^d/\Lambda$, $R$ commutes with $S_v$ if $Rv-v\in \Lambda$. We are interested in order 3 rotations on the hexagonal torus lattice $\Lambda = \Lambda_R(A_2)$, with order 3 shifts $v \in \tfrac{1}{3} \Lambda$. We can generically take, in the lattice basis
\begin{equation}\label{eq:genericHexagonal}
    R = \left(\begin{array}{cc}
         0 & -1 \\
        1 & -1 \\
        \end{array}\right) \qquad \text{and} \qquad v = \frac{1}{3} \left(\begin{array}{c} a \\ b \end{array}\right) \,,
\end{equation}
from which we can explicitly compute the commutator
\begin{equation}
    Rv-v = \frac{1}{3} \left(\begin{array}{c} -a-b \\ a-2b \end{array}\right),
\end{equation}
and conclude that the shifts that commute are in the weight lattice $\Lambda_W(A_2)$:
\begin{equation}\label{eq:shiftcomm3}
    R S_v R^{-1} S_v^{-1} = 1 \, \Longleftrightarrow \, a+b = 0 \text{ mod 3} \Longleftrightarrow \, v \in \Lambda_W(A_2) \subset \frac{1}{3} \Lambda_R(A_2)\,.
\end{equation}
This almost coincides with the condition for modular invariance \eqref{eq:modInv} for order 3 shifts:
\begin{equation}\label{eq:order3shift}
    (v^*)^2 \in \frac{2 \Z}{3} \Longleftrightarrow \, v^* \in \Lambda_W(A_2) \subset \frac{1}{3} \Lambda_R(A_2)\,.
\end{equation}
The only difference is that in \eqref{eq:order3shift} $v^*$ appears instead of $v$. We remind the reader that $v^*$ is the projection of $v$ on the lattice $I$ left invariant by the associated point group element. When  \eqref{eq:shiftcomm3} and \eqref{eq:order3shift} become equivalent, this means that the shifts permitted by modular invariance are exactly those that keep the orbifold Abelian, thus not solving the problem outlined at the beginning of this Section. 

We will now impose \eqref{eq:order3shift} for all cyclic subgroups of the point group. As the counting in \eqref{eq:primeProj} indicates, there are 4 such subgroups inside $\Z_3 \times \Z_3$, generated respectively by $f,g,fg$ and $fg^{-1}$.

First, one can try to make the orbifold non-commuting by adding a shift in a single $T^2$ component, either on the left movers or on the right movers. Without loss of generality, we can take $f$ to act as an order 3 rotation $R$. Then $g$ either acts as the identity, as $R$, or as $R^{-1}$. Using the freedom in redefining shifts on a torus as explained in \hyperref[app:shifts]{Appendix C}, we can put the shift entirely on $g$.
Then if $g$ acts as identity, one can see that any non-commuting shift will have to violate both \eqref{eq:shiftcomm3} and \eqref{eq:order3shift} since they coincide. This means the orbifold is not modular invariant anymore since it violates level matching. If $g$ acts as a non-trivial rotation, then naively $v$ does not enter in the level matching condition \eqref{eq:modInv}. This is true in the cyclic subgroup generated by $g$ since there is no invariant lattice, but either the subgroup generated by $fg$ or $fg^{-1}$ does have an invariant lattice, since the point group acts as identity there. Then if $v$ does not satisfy \eqref{eq:shiftcomm3} it will violate modular invariance in that sector. This shows that a shift vector in a single 2D component cannot do the job.

The next step is to try to combine shifts in different 2D blocks.
First, one could try to combine shifts on different $T^2$ components of a definite chirality, say all on the left. All vectors $v$ that are in $\tfrac{1}{3} \Lambda_R(A_2)$ but not in $\Lambda_W(A_2)$ satisfy 
\begin{equation}v^2 \in \frac{2}{9}+\frac{2 \Z}{3}.\label{ffw}\end{equation} If we insist on keeping a base $T^2$ on which the orbifold group acts by a pure geometric shift, then we have at most two components of $v$ (say, for $g$) satisfying \eq{ffw} where $f$ acts non-trivially, and summing their norms will not be enough to satisfy the modular invariance condition \eqref{eq:modInv}.

Up to now, we have excluded the case where all components of $v$ are all left (or right) movers. On the other hand, \eqref{eq:modInv} shows that a vector whose left and right components have the same norm satisfies level matching automatically.

%This is the case, for example, for geometric, left-right symmetric actions. But we can also use a non-geometric action, in which the shift acts on different components on the left and on the right. This is exactly our approach.

Not every point group admits such a $v$ however. To have left and right components of the invariant lattice cancel against each other, one would need to to have some element act as identity on both a $T^2$ on the left, and on a (possibly different) $T^2$ on the right. If we also insist on a having a base torus with a pure geometric shift action to get rid of massless twisted sectors, this means the twist vector is of the form $\Vec{\theta}=(0,0,\theta_3)$ both on the left and on the right. Lifting this through \eqref{eq:spinlift} one sees that this action would not preserve any local supercharge. Thus no such orbifold could satisfy condition \textbf{III.} of \hyperref[sec:groupth]{Section 3}.

Therefore, there is no modular invariant solution whose generators $f,g$ do not commute, and that acts purely as a geometric shift on a $T^2$ factor. This means we have to drop this last condition, and look for orbifolds that do not admit a decompactification limit. The only drawback is that we are not guaranteed to have lifted masses of all twisted sectors, so we will have to examine them one by one to make sure we have not reintroduced gravitini in the spectrum.

In \hyperref[sec:models]{Section 5} we explicitly present models where we realize precisely this mechanism. We use a shift which acts on different components on the left and on the right, but changes energy levels in exactly the same way on either side, thus preserving level matching. 

By this procedure we have extended the full orbifold space group to a non-Abelian group. Non-Abelian orbifolds may suffer from subtler global anomalies related to modular invariance. We will carefully examine them case by case and choose shifts that cancel those as well.

\subsection{$\Z_4 \times \Z_4$ point group}

We can repeat the analysis of the previous subsection for a $\Z_4 \times \Z_4$ point group, pretty much unchanged, since these actions also factorize nicely on $T^6\simeq T^2 \times T^2 \times T^2$.

Let us examine a 2D block individually: the lattice is the square one, $\Lambda = \Lambda_R(A_1 \times A_1)$. Taking a generic order 4 rotation and shift, in lattice coordinates
\begin{equation}\label{eq:z4generic}
    R = \left(\begin{array}{cc}
         0 & -1 \\
        1 & 0 \\
        \end{array}\right) \qquad \text{and} \qquad v = \frac{1}{4} \left(\begin{array}{c} a \\ b \end{array}\right) \,,
\end{equation}
we can explicitly the condition that they commute on $T^2/\Lambda$ to be
\begin{equation} \label{eq:z4explicit}
    RS_vR^{-1}S_v^{-1} = 1 \Longleftrightarrow \begin{cases}
        a = b = 0 \text{ mod 2} \\
        a = b \text{ mod 4}
    \end{cases}
\end{equation}
While the condition for modular invariance is
\begin{equation}
     (v^*)^2 \in \frac{2 \Z}{4} \Longleftrightarrow \, v^* \in \Lambda_W(A_1\times A_1) = \frac{1}{2} \Lambda_R(A_2)
\end{equation}
From which we see that also in this case any single 2D vector which is compatible with modular invariance necessarily commutes with rotations. The same considerations as the last Subsection lead again to the necessity to have the non-commuting shifts with components both on the left and on the right, such that level matching is not spoiled.

\section{The new models}\label{sec:models}

As shown in the previous two Sections, the only Abelian point groups that can realize the cancellation of the vacuum energy without SUSY are $\Z_2 \times \Z_2 \times \Z_2$, $\Z_3\times \Z_3$, $\Z_4 \times \Z_4$. Representing them as a vector as in \eqref{eq:f} and \eqref{eq:g}, we can run a full scan of the fermionic actions that realize the properties outlined at the beginning of \hyperref[sec:groupth]{Section 3}. From those we compute the corresponding bosonic actions in terms of twist vectors, and select for those twists that allow for the addition of shifts that get rid of the potentially problematic sectors in the partition function sum \eqref{eq:orbifold_partition_function}. As additional material to this paper, we attach a simple Python code which produces the full list of point group representations. The total count 240 $\Z_3\times \Z_3$ actions, and 1125 $\Z_4\times\Z_4$ actions. As mentioned earlier, there are only 2 possible $\Z_2 \times \Z_2 \times \Z_2$ actions, but in each of them one of the generators acts only as $(-1)^{F_{L/R}}$ at the level of the point group.

In this Section we present some models satisfying all the properties outlined in the previous Sections, fully explicitly.

\subsection{$\Z_3 \times \Z_3$} \label{sec:z3example}
Take as a fermionic action
\begin{equation}  \label{eq:z3SpinAction}
    \begin{aligned}
        f =& \, (1,1,\phi,\phi^2  |1,\phi,\phi,\phi)\,,\\
        g =& \, (\phi^2,\phi^2,\phi,\phi | \phi,\phi^2,1,1)\,.
    \end{aligned}
\end{equation}
where here and in the following, $\phi= e^{2 \pi i /3}$ is a third root of unity.
The generators preserve different supercharges, so SUSY is broken in the orbifolded theory. From the counting in \eqref{eq:projective} there are two more independent elements to check:
\begin{equation}
    \begin{aligned}
        fg =& \,(\phi^2,\phi^2,\phi^2,1  |\phi,1,\phi,\phi)\,, \\
        fg^2 =& \, (\phi,\phi,1,\phi | \phi^2,\phi^2,\phi,\phi)\,,
    \end{aligned}
\end{equation}
both of which leave some supercharge invariant.
\\To write the bosonic action, we apply the projection \eqref{eq:spinproj} to find the twist vectors: 
\begin{equation}\label{eq:bosZ3}
    \begin{aligned}
        f: \qquad \vec{\theta}^L = \left(0,\frac{1}{3},\frac{5}{3} \right), \qquad \vec{\theta}^R = \left(\frac{1}{3},\frac{1}{3},\frac{4}{3} \right), \\
        g: \qquad \vec{\theta}^L = \left(\frac{4}{3},0,0 \right), \qquad \vec{\theta}^R = \left(0,\frac{2}{3}, \frac{2}{3} \right).
    \end{aligned}
\end{equation}
The action factorizes as order 3 rotations on $T^2\times T^2 \times T^2$, so the lattice with this symmetry is
\begin{equation}
    \Lambda \simeq \Lambda_R(A_2) \oplus \Lambda_R(A_2) \oplus \Lambda_R(A_2)\,,
\end{equation}
where the radius is fixed at the self-dual value by the action of the orbifold, which includes T-duality elements like any asymmetric orbifold.

The two generators of course commute, and one can explicitly compute that
\begin{equation}
    \begin{aligned}
        \mathcal{Z}_L^F[f,g] &= 3 q^{\frac{1}{9}} + ... \,,\\
        \mathcal{Z}_R^F[f,g] &= e^{\frac{2}{9}i \pi } \left( 1 + e^{\frac{1}{3}i\pi} - 3q^{1/3}\right) + ...\,,
    \end{aligned}
\end{equation}
so we need to add shifts to get rid of it in the sum \eqref{eq:orbifold_partition_function}. A solution that avoids the anomalies discussed in \hyperref[sec:anomalies]{Section 6} is to take a shift in the invariant components of $g$ but not of $f$, such that the norm on the left and on the right is the same. In particular in each 2D block the invariant lattice is $I_2 \simeq \Lambda_R(A_2)$, with basis the two simple roots $\alpha_1,\alpha_2$. We add a shift
\begin{equation}\label{eq:shiftz3}
    v_2 = \frac{1}{3} \alpha_1
\end{equation}
to the action of $g$ in the two copies of $I_2$ associated to $X_3^L,X_4^L$ on the left, and $X_1^R,X_2^R$ on the right. What appears in the anomaly cancellation condition on the cyclic subgroup generated by $g$ is the combination
\begin{equation}
    (v^*)^2 = (v_2)^2-(v_2)^2= 0\,,
\end{equation}
so we have not introduced any global anomalies by adding this shift. Equivalently, we have not spoiled level matching \eqref{eq:levelMatch} by adding the same energy shift on the left and on the right, similarly to how a geometric shift would act.

The group generated by $f$ and $g$ with these choices of shifts has order 27, and it is traditionally called $\Delta_{27}$ in the physics literature.\footnote{ \label{footnote1}In addition to being known in the literature as $\Delta_{27}$, it is isomorphic to the Heisenberg group over the finite field $\Z_3$. A typical presentation is given by  $$\langle x,y,z | x^3 = y^3 = z^3 = 1, x y x^{-1}y^{-1}=z, xz = zx, yz = zy \rangle \,.$$ Our presentation is already in this form, by identifying $x = f, y = g$.}
The analysis of \hyperref[sec:anomalies]{Section 6} shows that this level matching condition is not enough to guarantee modular invariance at all loops, because of an extra potential anomaly that is not detected at one-loop and is carried by the non-Abelian representation of $\Delta_{27}$, which we call $V_{NA}$.

Since this is a $\mathbb{Z}_3$ anomaly, our task is  to show that with our choice of shift the net number of copies of $V_{NA}$ appearing in the matter fields vanishes modulo 3. Fermions only see the point group, and therefore transform in an Abelian representation. However, the 2D chiral scalars do transform in a non-Abelian representation of $\Delta_{27}$, and may carry copies of $V_{NA}$. Through fermionization we can precisely compute the representation and show it is not anomalous.

%Intuitively, since we chose essentially the same shifts on the left and on the right, we have the same number of left-mover and right-mover copies of $\rho_{NA}$, so the anomaly cancels out. This is automatically the case for fermions, since the shift does not act on them. What is modified is the representation in which 2D chiral scalars transform, and where the potential anomaly may hide. Through fermionization we can precisely compute the representation and show it is not anomalous.

The fermionization map for chiral bosons on the $A_2$ lattice maps their CFT isomorphically to a theory of three free fermions in the fundamental of $SU(3)$ \cite{Witten:1983ar}. To build this map we need an expression for the three fundamental weights of $A_2$ in terms of the roots:
\begin{equation} \label{eq:weights}
    \begin{aligned}
         w_1 &= \frac{2}{3}\alpha_1 + \frac{1}{3} \alpha_2, \quad w_2 &= -\frac{1}{3}\alpha_1 + \frac{1}{3} \alpha_2 , \quad w_3 &= -\frac{1}{3}\alpha_1 - \frac{2}{3} \alpha_2\,.
    \end{aligned}
\end{equation}
Then we build fermions out of the chiral bosons $X^{L/R}$ as 
\begin{equation}\label{eq:fermionize}
    \psi_k^{L/R} \sim :e^{i w_k \cdot X^{L/R} }\,.
\end{equation}
From this map, together with the internal products $w_i \cdot \alpha_j$, we can recover the induced action on the $\psi$'s from the explicit expression of rotations and shifts as in \eqref{eq:genericHexagonal} to be
\begin{equation}\label{eq:fermionizemaps}
    \begin{aligned}
        R:\left(
            \begin{array}{c}
            \psi_1 \\
            \psi_2 \\
            \psi_3
            \end{array}
            \right)
            &\mapsto
            \left(
            \begin{array}{ccc}
            0 & 1 & 0 \\
            0 & 0 & 1 \\
            1 & 0 & 0
            \end{array}
            \right)
            \left(
            \begin{array}{c}
            \psi_1 \\
            \psi_2 \\
            \psi_3
            \end{array}
            \right)\,, \\
        S_{v}:
               \left(
                \begin{array}{c}\psi_1 \\
                \psi_2 \\
                \psi_3
                \end{array}
                \right)
                &\mapsto
                \left(
                \begin{array}{ccc}
                \phi^a & 0 & 0 \\
                0 & \phi^{b-a} & 0 \\
                0 & 0 & \phi^{-b}
                \end{array}
                \right)
                \left(
                \begin{array}{c}
                \psi_1 \\
                \psi_2 \\
                \psi_3
                \end{array}
            \right)\,.
    \end{aligned}
\end{equation}
%So for our choice of shift \eqref{eq:shiftz3}, the action is
%\begin{equation}
%    S_{v_2} = \left(
%                \begin{array}{ccc}
 %               \phi & 0 & 0 \\
%                0 & \phi^{2} & 0 \\
%                0 & 0 & 1
%                \end{array}
%                \right)\,.
%\end{equation}
%This matrix, together with the permutation in \eqref{eq:fermionizemaps}, defines the unique\footnote{Up to complex conjugation. The anomaly is associated to the first Pontryagin class of the representation \cite{Freed:1987qk}, which depends only on the underlying real representation. For these purposes, we can thus treat $\Delta_{27}$ as having a unique faithful representation.} faithful representation of $\Delta_{27}$. Let us call this 3D representation $V_3$.
%The other representations that appear on chiral bosons have no shifts, so they factor through the point group $\Z_3 \times \Z_3$. Since they are Abelian, they are 1D, and we label them by the character associated to their generators: call $A_{m,n}$ the representation which maps $(f,g)$ to $(\phi^m,\phi^n)$.

By carefully applying \eqref{eq:fermionize} and \eqref{eq:fermionizemaps} to all coordinates, we find the total representation to be, on the left and right, respectively:

\begin{equation}
    \begin{aligned}
        V^L &= 2\, A_{0,0} \oplus A_{0,1} \oplus A_{0,2} \oplus A_{1,0} \oplus A_{2,0} \oplus V_{NA}\\
        V^R &= 2\, A_{0,0} \oplus 2\, A_{1,2} \oplus 2 \,A_{2,1} \oplus V_{NA}
    \end{aligned}
\end{equation}
where $A_{m,n}$ denotes the (Abelian) representation which maps $(f,g)$ to $(\phi^m,\phi^n)$. 
We see that the truly non-Abelian part is the same on the left and on the right, so there is no extra anomalies. The Abelian part is level matched, and therefore also anomaly-free.

Finally, since we do not have a large torus to make the orbifold freely acting, we expect massless twisted sectors. We thus need to examine these sectors in detail to make sure we have not accidentally restored SUSY by introducing twisted gravitini.

Checking this is a direct application of the discussion at the end of \hyperref[sec:asymmOrb]{Section 2}. Recall that to have a massless vector in the $h$-twisted sector one needs the twist $\theta_h$ to be trivial either on the left or on the right. The point group generated by \eqref{eq:bosZ3} has no non-trivial element with this property and is Abelian, so all that remains to check is the commutator subgroup $[\Delta_{27},\Delta_{27}]\simeq \Z_3$, which by construction is given by pure translations.
Recall that shift that arise in commutators are of the form:
\begin{equation}
    \Tilde{v} = Rv-v = \frac{1}{3} \left(\begin{array}{c} -a-b \\ a-2b \end{array}\right) \in \Lambda_W(A_2) \,.
\end{equation}
Since the twist is trivial both on the left and on the right, the zero point energy contributions vanish on either side, and the ground state energies \eqref{eq:Hvec},\eqref{eq:Hspinor} for both a vector and a spinor are
\begin{equation}\label{eq:GSeasy}
    \begin{aligned}
        H^0_L = \min_{(p_L,p_R) \in I^*} \frac{(v_L-p_L)^2}{2} \,, \\
        H^0_R = \min_{(p_L,p_R) \in I^*} \frac{(v_R-p_R)^2}{2} \,,
    \end{aligned}
\end{equation}
where for a trivial twist
\begin{equation}
    I^* = I = \Gamma^{2,2}(A_2) \,.
\end{equation}
Let us specialize to the choice of shift described above. Recall that we put the left and right moving shifts on different $T^2$ components, so for example on the first $T^2$, the commutator subgroup acts with a shift
\begin{equation}
    (v_L,v_R) = (0,Rv_2-v_2) \in \{0\}\times \Lambda_W(A_2) \,,
\end{equation}
which, crucially, is not in $I^*$. From the form \eqref{eq:extraSymm} of $\Gamma^{2,2}$ one can explicitly see that it is impossible to set both the left and right moving Hamiltonian to zero in \eqref{eq:GSeasy}, since $p_L=0$ implies $p_R \in \Lambda_R(A_2)$, but $v_R \notin \Lambda_R(A_2)$. Conversely, setting the right moving Hamiltonian to zero requires $p_R \in \left\{\Lambda_W(A_2) \setminus \Lambda_R(A_2)\right\}$, which implies $p_L \neq 0$.
This means there are no massless gravitini, so SUSY is indeed broken in the full orbifold theory, and this model realizes the mechanism of \hyperref[sec:groupth]{Section 3}.
\\
To summarize, we provide the full action of the orbifold generators on bosons, in the $\Lambda_R(A_2)^{\oplus 3}$ lattice basis: $f$ acts as
\begin{equation}
    f:\left\{
    \begin{aligned}
        \left(
        \begin{array}{c}
             X_1^L \\
             X_2^L
        \end{array}
        \right)
        &\longmapsto
        \left(
        \begin{array}{c}
             X_1^L \\
             X_2^L
        \end{array}
        \right) \,,  & \quad
        \left(
        \begin{array}{c}
             X_1^R \\
             X_2^R
        \end{array}
        \right)
        &\longmapsto
        \left(
        \begin{array}{cc}
         0 & -1 \\
         1 & -1 \\
        \end{array}
        \right)
        \left(
        \begin{array}{c}
             X_1^R \\
             X_2^R
        \end{array}
        \right)\,, \\ %end line here    
        \left(
        \begin{array}{c}
             X_3^L \\
             X_4^L
        \end{array}
        \right)
        &\longmapsto
        \left(
        \begin{array}{cc}
         0 & -1 \\
         1 & -1 \\
        \end{array}
        \right)
        \left(
        \begin{array}{c}
             X_3^L \\
             X_4^L
        \end{array}
        \right) \,, & \quad
        \left(
        \begin{array}{c}
             X_3^R \\
             X_4^R
        \end{array}
        \right)
        &\longmapsto
        \left(
        \begin{array}{cc}
         0 & -1 \\
         1 & -1 \\
        \end{array}
        \right)
        \left(
        \begin{array}{c}
             X_3^R \\
             X_4^R
        \end{array}
        \right) \,,\\ %end line here    
        \left(
        \begin{array}{c}
             X_5^L \\
             X_6^L
        \end{array}
        \right)
        &\longmapsto
        \left(
        \begin{array}{cc}
         -1 & 1 \\
         -1 & 0 \\
        \end{array}
        \right)
        \left(\begin{array}{c}
                X_5^L \\
                X_6^L
              \end{array}\right)
        ,  & \quad
        \left(
        \begin{array}{c}
             X_5^R \\
             X_6^R
        \end{array}
        \right)
        &\longmapsto
        \left(
        \begin{array}{cc}
         0 & -1 \\
         1 & -1 \\
        \end{array}
        \right)
        \left(\begin{array}{c}
                X_5^R \\
                X_6^R
              \end{array}\right)
        ,
    \end{aligned}
    \right.
\end{equation}
with an extra $(-1)^F$ on all space time fermions.
While $g$ acts as
\begin{equation}
    g:\left\{
    \begin{aligned}
        \left(
        \begin{array}{c}
             X_1^L \\
             X_2^L
        \end{array}
        \right)
        &\longmapsto
        \left(
        \begin{array}{cc}
         0 & -1 \\
         1 & -1 \\
        \end{array}
        \right)
        \left(
        \begin{array}{c}
             X_1^L \\
             X_2^L
        \end{array}
        \right) \,,  & \quad
        \left(
        \begin{array}{c}
             X_1^R \\
             X_2^R
        \end{array}
        \right)
        &\longmapsto
        \left(
        \begin{array}{c}
             X_1^R \\
             X_2^R
        \end{array}
        \right)
        +\left(\begin{array}{c}
                1/3 \\
                0
              \end{array}\right)\,, \\ %end line here    
        \left(
        \begin{array}{c}
             X_3^L \\
             X_4^L
        \end{array}
        \right)
        &\longmapsto
        \left(
        \begin{array}{c}
             X_3^L \\
             X_4^L
        \end{array}
        \right)
        +\left(\begin{array}{c}
                1/3 \\
                0
              \end{array}\right) \,,  & \quad
        \left(
        \begin{array}{c}
             X_3^R \\
             X_4^R
        \end{array}
        \right)
        &\longmapsto
        \left(
        \begin{array}{cc}
         -1 & 1 \\
         -1 & 0 \\
        \end{array}
        \right)
        \left(
        \begin{array}{c}
             X_3^R \\
             X_4^R
        \end{array}
        \right) \,,\\ %end line here    
        \left(
        \begin{array}{c}
             X_5^L \\
             X_6^L
        \end{array}
        \right)
        &\longmapsto
        \left(\begin{array}{c}
                X_5^L \\
                X_6^L
              \end{array}\right)
        \,,  & \quad
        \left(
        \begin{array}{c}
             X_5^R \\
             X_6^R
        \end{array}
        \right)
        &\longmapsto
        \left(
        \begin{array}{cc}
         -1 & 1 \\
         -1 & 0 \\
        \end{array}
        \right)
        \left(\begin{array}{c}
                X_5^R \\
                X_6^R
              \end{array}\right)\,,
    \end{aligned}
    \right.
\end{equation}
with an extra $(-1)^{F_L}$ on all left moving spacetime fermions.

\subsection{$\Z_4 \times \Z_4$} \label{sec:z4example}

We start with the fermionic action
\begin{equation}  \label{eq:z4SpinAction}
    \begin{aligned}
        f =& \, (1,1,i,-i|i,1,i,-1)\,,\\
        g =& \, (i,-1,-i,-1 |-1,-1,1,1)\,.
    \end{aligned}
\end{equation}
These preserve different supercharges, so SUSY is explicitly broken in the untwisted sector. The other indepedent elements one needs to check in this case to ensure local SUSY conservation are $fg,fg^{-1},fg^2,f^2g$, and they all explicitly preserve supercharges.

The bosonic projection leads to twist vectors
\begin{equation}\label{eq:bosZ4}
    \begin{aligned}
        f: \qquad \vec{\theta}^L = \left(0,\frac{1}{4},\frac{7}{4} \right), \qquad \vec{\theta}^R = \left(\frac{1}{4},\frac{1}{2},\frac{7}{4} \right), \\
        g: \qquad \vec{\theta}^L = \left(\frac{3}{4},0,\frac{7}{4} \right), \qquad \vec{\theta}^R = \left(0,\frac{1}{2}, \frac{1}{2} \right).
    \end{aligned}
\end{equation}
The action factorizes as order 4 rotations on $T^2 \times T^2 \times T^2$, so the lattice with this symmetry is the square one:
\begin{equation}
    \Lambda = \Lambda_R(A_1\times A_1)^{\oplus 3}\,,
\end{equation}
with the radius fixed at the self-dual value by the symmetries of the orbifold.
The generators commute at the level of the point group, and the mixed sector is non-vanishing:
\begin{equation}
    \begin{aligned}
        \mathcal{Z}_L^F[f,g] &\sim 2(1-i) \, \, q^{\frac{7}{48}}+ ...\,, \\ \mathcal{Z}_R^F[f,g] &\sim -2  q^{\frac{1}{48}} + ...\,.
    \end{aligned}
\end{equation}
We get rid of this sector with the same strategy of the previous subsection: we add a shift to $g$ in its invariant components, in particular we can choose
\begin{equation}\label{eq:z4shift}
    v_2 = \frac{1}{4}\alpha
\end{equation}
in the $X^1_R$ direction on the the right, and on the $X^3_L$ on the left, where $\alpha$ is the simple root of the $A_1$ sublattice associated to either coordinate.
\\With this choice of shift, the group generated by $f$ and $g$ is a semidirect product $(\Z_4 \times \Z_4) \rtimes \Z_4$\footnote{One can see the $\Z_4 \times \Z_4$ factor as representing order 4 translations on $\R^2/\Z^2$ in each coordinate direction, and the rotation acts on them by conjugating one into the other. A presentation is given by $$ \langle R,s_1,s_2 \, | \, R^4 = s_1^4 = s_2^4 = (Rs_1)^4 = (R s_2)^4 = 1 , Rs_1R^{-1} = s_2, s_1 s_2 = s_2 s_1 \rangle$$
Our generators are already in this form, with an isomorphism given by $R=f, s_1 = g$.\label{fn:Z2xZ2xZ2xZ4}}.

The Abelian anomaly is certainly canceled, because the norm of the shift is the same on the left and on the right and thus cannot spoil level matching. It is considerably harder to compute the relevant bordism group to determine whether there can be additional global anomalies, since the group has even order and the corresponding spectral sequence has many more potentially non-trivial differentials. Nevertheless, the same consideration of the previous subsection applies: our choice of shift enhances a 2D Abelian representation to the same non-Abelian representation of the space group on the left and on the right. Therefore the only part of the representation which is truly non-Abelian is not chiral, and there is no global anomaly.
Let us show this in detail through fermionization. For chiral bosons living on the square lattice, the fermionization map is simpler:
\begin{equation}
    \psi_k^{L/R} \sim :e^{iX_k^{L/R}}:
\end{equation}
We can write the maps induced from \eqref{eq:z4explicit} in complex coordinates $\psi_k,\overline{\psi}_k$ as
\begin{equation}
    R: \left(
        \begin{array}{c}
         \psi_1  \\
         \overline{\psi}_1  \\
         \psi_2  \\
         \overline{\psi}_2  \\
        \end{array}
        \right)
        \mapsto\left(
        \begin{array}{cccc}
         0 & 0 & 1 & 0 \\
         0 & 0 & 0 & 1 \\
         0 & 1 & 0 & 0 \\
         1 & 0 & 0 & 0 \\
        \end{array}
        \right)
        \left(
        \begin{array}{c}
         \psi_1  \\
         \overline{\psi}_1  \\
         \psi_2  \\
         \overline{\psi}_2  \\
        \end{array}
        \right), \qquad
        S: \left(
        \begin{array}{c}
         \psi_1  \\
         \overline{\psi}_1  \\
         \psi_2  \\
         \overline{\psi}_2  \\
        \end{array}
        \right) \mapsto
        \left(
        \begin{array}{cccc}
         i^a & 0 & 0 & 0 \\
         0 & i^{-a} & 0 & 0 \\
         0 & 0 & i^b & 0 \\
         0 & 0 & 0 & i^{-b} \\
        \end{array}
        \right)
        \left(
        \begin{array}{c}
         \psi_1  \\
         \overline{\psi}_1  \\
         \psi_2  \\
         \overline{\psi}_2  \\
        \end{array}
        \right)
\end{equation}
When $a=0,b=1$, these two matrices generate a faithful real 4D representation of the space group $(\Z_4 \times \Z_4) \rtimes \Z_4$, let us call it $V_{NA}$. In addition, similarly to the last Section, we denote Abelian representations induced by $\Z_4\times \Z_4$ as $A_{m,n}$ when they send $(f,g)$ to $(i^m,i^n)$. The total representation is then
\begin{equation}
    \begin{aligned}
        V^L &= 2\, A_{0,0} \oplus A_{0,1} \oplus A_{0,2} \oplus A_{0,3} \oplus A_{1,1} \oplus A_{2,2} \oplus A_{3,3} \oplus V_{NA}\\
        V^R &= 5\, A_{2,2} \oplus A_{0,2} \oplus A_{1,2} \oplus A_{3,2}  \oplus V_{NA}
    \end{aligned}
\end{equation}
As expected, the truly non-Abelian subrepresentation is non-chiral, so it does not contribute extra anomalies.

Lastly, we need to look for potential gravitini in the twisted sectors. The point group never acts trivially either on the left or on the right, so we need only look in the commutator subgroup, which is pure shift.

The situation is similar to the previous Section: with the generic choices as in \eqref{eq:z4explicit} we compute the generic commutator in a 2D block
\begin{equation}
    Rv-v = \frac{1}{4} \left(\begin{array}{c} -a-b \\ a-b \end{array}\right)\,.
\end{equation}
This is clearly not in $\Lambda_R(A_1\times A_1)$ for generic values of $a,b$, and in particular it is not for $a=1,b=0$ as in \eqref{eq:z4shift}. This means that, e.g, a shift $v_2$ on the left and nothing on the right in the same 2D component will never give a massless state through \eqref{eq:GSeasy} precisely for the same reason as in the $\Z_3 \times \Z_3$ example. Our choice \eqref{eq:z4explicit} is exactly of this form, and will therefore lift the masses of any potential twisted gravitino.

In summary, we take the action on bosons to be for $f$:
\begin{equation}
    f:\left\{
    \begin{aligned}
        \left(
        \begin{array}{c}
             X_1^L \\
             X_2^L
        \end{array}
        \right)
        &\longmapsto
        \left(
        \begin{array}{c}
             X_1^L \\
             X_2^L
        \end{array}
        \right) \,,  & \quad
        \left(
        \begin{array}{c}
             X_1^R \\
             X_2^R
        \end{array}
        \right)
        &\longmapsto
        \left(
        \begin{array}{cc}
         0 & -1 \\
         1 & 0 \\
        \end{array}
        \right)
        \left(
        \begin{array}{c}
             X_1^R \\
             X_2^R
        \end{array}
        \right)\,, \\ %end line here    
        \left(
        \begin{array}{c}
             X_3^L \\
             X_4^L
        \end{array}
        \right)
        &\longmapsto
        \left(
        \begin{array}{cc}
         0 & -1 \\
         1 & 0 \\
        \end{array}
        \right)
        \left(
        \begin{array}{c}
             X_3^L \\
             X_4^L
        \end{array}
        \right) \,, & \quad
        \left(
        \begin{array}{c}
             X_3^R \\
             X_4^R
        \end{array}
        \right)
        &\longmapsto
        \left(
        \begin{array}{c}
             -X_3^R \\
             -X_4^R
        \end{array}
        \right) \,,\\ %end line here    
        \left(
        \begin{array}{c}
             X_5^L \\
             X_6^L
        \end{array}
        \right)
        &\longmapsto
        \left(
        \begin{array}{cc}
         0 & 1 \\
         -1 & 0 \\
        \end{array}
        \right)
        \left(\begin{array}{c}
                X_5^L \\
                X_6^L
              \end{array}\right)
        ,  & \quad
        \left(
        \begin{array}{c}
             X_5^R \\
             X_6^R
        \end{array}
        \right)
        &\longmapsto
        \left(
        \begin{array}{cc}
         0 & 1 \\
         -1 & 0 \\
        \end{array}
        \right)
        \left(\begin{array}{c}
                X_5^R \\
                X_6^R
              \end{array}\right)
        ,
    \end{aligned}
    \right.
\end{equation}
with an extra $(-1)^{F}$ on all spacetime fermions, and for $g$:
\begin{equation}
    g:\left\{
    \begin{aligned}
        \left(
        \begin{array}{c}
             X_1^L \\
             X_2^L
        \end{array}
        \right)
        &\longmapsto
        \left(
        \begin{array}{cc}
         0 & 1 \\
         -1 & 0 \\
        \end{array}
        \right)
        \left(
        \begin{array}{c}
             X_1^L \\
             X_2^L
        \end{array}
        \right) \,,  & \quad
        \left(
        \begin{array}{c}
             X_1^R \\
             X_2^R
        \end{array}
        \right)
        &\longmapsto
        \left(
        \begin{array}{c}
             X_1^R \\
             X_2^R
        \end{array}
        \right)+
        \left(
        \begin{array}{c}
             1/4 \\
             0
        \end{array}
        \right)\, ,\\ %end line here    
        \left(
        \begin{array}{c}
             X_3^L \\
             X_4^L
        \end{array}
        \right)
        &\longmapsto
        \left(
        \begin{array}{c}
             X_3^L \\
             X_4^L
        \end{array}
        \right) + \left(
        \begin{array}{c}
             1/4 \\
             0
        \end{array}
        \right)
        \,, & \quad
        \left(
        \begin{array}{c}
             X_3^R \\
             X_4^R
        \end{array}
        \right)
        &\longmapsto
        \left(
        \begin{array}{c}
             -X_3^R \\
             -X_4^R
        \end{array}
        \right) \,,\\ %end line here    
        \left(
        \begin{array}{c}
             X_5^L \\
             X_6^L
        \end{array}
        \right)
        &\longmapsto
        \left(
        \begin{array}{cc}
         0 & 1 \\
         -1 & 0 \\
        \end{array}
        \right)
        \left(\begin{array}{c}
                X_5^L \\
                X_6^L
              \end{array}\right)
        ,  & \quad
        \left(
        \begin{array}{c}
             X_5^R \\
             X_6^R
        \end{array}
        \right)
        &\longmapsto
        \left(\begin{array}{c}
                -X_5^R \\
                -X_6^R
              \end{array}\right)
        .
    \end{aligned}
    \right.
\end{equation}
with a $(-1)^{F_L}$ acting on left-moving spacetime fermions.

\subsection{$S_3 \times \Z_3$} \label{sec:S3example}
We can also find some examples of models satisfying our properties with an orbifold group which is non-Abelian already at the point group level.
A classification of these models is considerably harder, since the tools of \hyperref[sec:groupth]{Section 3} do not apply. On the other hand, once a given point group is found, since many elements already do not commute, there may be no need to engineer shifts to get rid of potentially non-vanishing sectors in the partition function.

As an example, we can choose the fermionic action of our generators to be:
\begin{equation}\label{eq:nonAbPoint}
    \begin{aligned}
        f &= \, (\phi,\phi^2,\phi,\phi^2  |1,1,\phi^2,\phi), \\
         g & =  ( \left(
                        \begin{array}{cccc}
                         0 & 0 & 0 & 1 \\
                         0 & 0 & 1 & 0 \\
                         0 & 1 & 0 & 0 \\
                         1 & 0 & 0 & 0 \\
                        \end{array}
                    \right)| -1,-1,1,1),
    \end{aligned}
\end{equation}
where $\phi=e^{2 \pi i/3}$ is a third root of unity, and we have written down explicitly the non-diagonal part of the generator. The rest of the notation is borrowed from the previous Sections. The two generators preserve different supercharges, so SUSY is broken. There are more independent elements to check in the order 18 group that $f$ and $g$ generate, but all of them explicitly preserve at least a supercharge, so $\mathcal{Z}[1,h]$ vanishes for all $h \in G$.

The abstract group acting faithfully on fermions is $S_3\times \Z_3$\footnote{The permutation group $S_3$ can be presented as $$\langle \, \sigma_{12}, \sigma_{123} | \, \sigma_{12}^2 = \sigma_{123}^3 = 1, \, \sigma_{12}\sigma_{123}\sigma_{12} = \sigma_{123}^{-1} \, \rangle$$
The map to our generators is given by $\sigma_{12} = g, \sigma_{123} = f^{-1}g^{-1}fg$. The $\Z_3$ factor is generated by $ c = fgfg$, which can be checked to be order 3 and central.\label{fn:S3xZ3}}.
Since the fermionic action is not diagonal anymore, to get the bosonic actions we need to explicitly use \eqref{eq:spinproj}. In an appropriate basis, the result is
\begin{equation}\label{eq:bosS3}
    \begin{aligned}
        f_L &= \left(\begin{array}{ccc}
                R(4 \pi/3) & 0 & 0 \\
                 0 & I_2 & 0 \\
                 0 & 0 & I_2 \\
                \end{array}\right), \qquad
        &f_R =& \left(\begin{array}{ccc}
                R(4 \pi/3) & 0 & 0 \\
                 0 & R(2 \pi /3) & 0 \\
                 0 & 0 & I_2 \\
                \end{array}\right), \\
        g_L &= \left(\begin{array}{ccc}
                \mathcal{P} & 0 & 0 \\
                 0 & -I_2 & 0 \\
                 0 & 0 & \mathcal{P} \\
                \end{array}\right), \qquad
        &g_R =& \left(\begin{array}{ccc}
                -I_2 & 0 & 0 \\
                 0 & -I_2 & 0 \\
                 0 & 0 & I_2 \\
                \end{array}\right), \\
    \end{aligned}
\end{equation}
where $R(\alpha)$ is a real rotation matrix, $I_2$ is the 2D identity, and $\mathcal{P}=$ diag$(-1,1)$ is a reflection along an axis.

A lattice with this symmetry is given by
\begin{equation}\label{eq:LatticeS3Z3}
    \Lambda_R(A_2) \oplus \Lambda_R(A_2) \oplus \Lambda_R(A_1) \oplus \Lambda^1_b.
\end{equation}
Where $\Lambda_b^1 \simeq l_b \, \Z$ gives a circle of arbitrary length.
Only the last direction is arbitrary, since nothing acts asymmetrically on it, while everything else is fixed at the self-dual radius.
The associated Narain lattice is then
\begin{equation}
    \Gamma^{6,6} = \Gamma^{2,2}(A_2) \oplus \Gamma^{2,2}(A_2) \oplus \Gamma^{1,1}(A_1) \oplus \Gamma_b^{1,1}.
\end{equation}
Since we do have a base circle in this case, we can put shifts on it for both generators, of order 3 for $f$ and of order 2 for $g$, to mass up almost all twisted sectors.
The only potentially dangerous twisted sectors are those related to the commutator subgroup, since the action on the base circle is necessarily trivial there.
For our group:
\begin{equation}
    [G,G] \simeq \Z_3 \simeq \langle f g f^{-1} g^{-1} \rangle
\end{equation}
so we have only have one twisted sector to check.
The corresponding twist vector is
\begin{equation} \label{eq:commutator_twist}
    fgf^{-1}g^{-1}: \qquad \vec{\theta}_L = \left(\frac{4}{3},0,0\right), \qquad \vec{\theta}_R = (0,0,0)
\end{equation}
Since it acts trivially on the right, through \eqref{eq:Hvec} and \eqref{eq:Hspinor} we can see that there is a massless gravitino in the twisted sector, which means that this model, as currently written, is actually supersymmetric.

We can however choose a shift vector that lifts this mass, as follows: we add to $f$ a shift
\begin{equation} \label{eq:shiftS3}
    v_2 = \frac{1}{3} \alpha_1 + \frac{2}{3} \alpha_2
\end{equation}
in the invariant component associated to $X_L^3,X_L^4$, which is a copy of $\Lambda_R(A_2)$. It satisfies \eqref{eq:order3shift}, so modular invariance is not spoiled, and it can be explicitly computed that the induced shift on the commutator $f g f^{-1}g^{-1}$ is
\begin{equation}
    \Tilde{v}_2 = \frac{2}{3} \alpha_1 + \frac{1}{3} \alpha_2
\end{equation}
purely on the left. With the twist \eqref{eq:commutator_twist} the ground state energies in the Ramond sector are the same as in \eqref{eq:GSeasy}, so for the same reason as in that Section, a shift which is in $\Lambda_W(A_2)\times\{0\}$ cannot produce a level matched massless gravitino.
This means that there is no SUSY restoration, and this model realizes the local SUSY mechanism with a non-Abelian point group. One might worry that the addition of the shift would enhance the space group to a bigger non-Abelian group, similar to what happened in the case of Abelian point groups. This can be checked explicitly not to happen: the addition of \eqref{eq:shiftS3} preserves the $S_3 \times \Z_3$ group structure also at the space group level.
We verify in \hyperref[sec:anomalies]{Section 6} that this group cannot have global anomalies other than those induced from Abelian subgroups, so we do not have to worry about additional modular invariance constraints other than level matching.

To summarize, as usual, we provide the explicit bosonic action:
$f$ acts as
\begin{equation}
    f:\left\{
    \begin{aligned}
        \left(
        \begin{array}{c}
             X_1^L \\
             X_2^L
        \end{array}
        \right)
        &\longmapsto
        \left(
        \begin{array}{cc}
         -1 & 1 \\
         -1 & 0 \\
        \end{array}
        \right)
        \left(
        \begin{array}{c}
             X_1^L \\
             X_2^L
        \end{array}
        \right) \,,  & \quad
        \left(
        \begin{array}{c}
             X_1^R \\
             X_2^R
        \end{array}
        \right)
        &\longmapsto
        \left(
        \begin{array}{cc}
         -1 & 1 \\
         -1 & 0 \\
        \end{array}
        \right)
        \left(
        \begin{array}{c}
             X_1^R \\
             X_2^R
        \end{array}
        \right)\,, \\ %end line here    
        \left(
        \begin{array}{c}
             X_3^L \\
             X_4^L
        \end{array}
        \right)
        &\longmapsto
        \left(
        \begin{array}{c}
             X_3^L \\
             X_4^L
        \end{array}
        \right)
        +\left(\begin{array}{c}
                1/3 \\
                2/3
              \end{array}\right)\,, & \quad
        \left(
        \begin{array}{c}
             X_3^R \\
             X_4^R
        \end{array}
        \right)
        &\longmapsto
        \left(
        \begin{array}{cc}
         0 & -1 \\
         1 & -1 \\
        \end{array}
        \right)
        \left(
        \begin{array}{c}
             X_3^R \\
             X_4^R
        \end{array}
        \right) \,,\\ %end line here    
        \left(
        \begin{array}{c}
             X_5^L \\
             X_6^L
        \end{array}
        \right)
        &\longmapsto
        \left(\begin{array}{c}
                X_5^L \\
                X_6^L
              \end{array}\right)
        +\left(\begin{array}{c}
                0 \\
                l_b/3
              \end{array}\right) \,,  & \quad
        \left(
        \begin{array}{c}
             X_5^R \\
             X_6^R
        \end{array}
        \right)
        &\longmapsto
        \left(\begin{array}{c}
                X_5^R \\
                X_6^R
              \end{array}\right)
        +\left(\begin{array}{c}
                0 \\
                l_b/3
              \end{array}\right) \,,
    \end{aligned}
    \right.
\end{equation}
with an extra $(-1)^{F_R}$ on right moving spacetime fermions.

Meanwhile, $g$ acts as
\begin{equation} \label{eq:gS3boson}
    g:\left\{
    \begin{aligned}
        \left(
        \begin{array}{c}
             X_1^L \\
             X_2^L
        \end{array}
        \right)
        &\longmapsto
        \left(
        \begin{array}{cc}
         -1 & 1 \\
         0 & 1 \\
        \end{array}
        \right)
        \left(
        \begin{array}{c}
             X_1^L \\
             X_2^L
        \end{array}
        \right),  & \quad
        \left(
        \begin{array}{c}
             X_1^R \\
             X_2^R
        \end{array}
        \right)
        &\longmapsto
        \left(
        \begin{array}{c}
            -X_1^R \\
            -X_2^R
        \end{array}
        \right)\,, \\ %end line here    
        \left(
        \begin{array}{c}
             X_3^L \\
             X_4^L
        \end{array}
        \right)
        &\longmapsto
        \left(
        \begin{array}{c}
             -X_3^L \\
             -X_4^L
        \end{array}
        \right)
        \,, & \quad
        \left(
        \begin{array}{c}
             X_3^R \\
             X_4^R
        \end{array}
        \right)
        &\longmapsto
        \left(
        \begin{array}{c}
             -X_3^R \\
             -X_4^R
        \end{array}
        \right) \,,\\ %end line here    
        \left(
        \begin{array}{c}
             X_5^L \\
             X_6^L
        \end{array}
        \right)
        &\longmapsto
        \left(\begin{array}{c}
                -X_5^L \\
                X_6^L
              \end{array}\right)
        +\left(\begin{array}{c}
                0 \\
                l_b/2
              \end{array}\right)
        \,,  & \quad
        \left(
        \begin{array}{c}
             X_5^R \\
             X_6^R
        \end{array}
        \right)
        &\longmapsto
        \left(\begin{array}{c}
                X_5^R \\
                X_6^R
              \end{array}\right)
        +\left(\begin{array}{c}
                0 \\
                l_b/2
              \end{array}\right) \,.
    \end{aligned}
    \right.
\end{equation}

\subsection{$D_6$} \label{sec:d6}

As a last example, take $f$ to act on fermions as
\begin{equation} \label{eq:fd6}
    f = (\zeta,\zeta^{-1},\zeta,\zeta^{-1} | 1,1,-1,-1) \quad \text{for}  \quad\zeta^6=1\,,
\end{equation}
and $g$ as in \eqref{eq:nonAbPoint}.

Again it is clear that the two generators do not commute, and preserve different supercharges, so SUSY is broken. The abstract group they generate is the dihedral group of order 12, $D_6$\footnote{\label{fn:D_6_presentation}The dihedral group $D_n$, of order $2n$, has a canonical presentation as the symmetry group of an $n$-gon $$\langle s, r \, | \, s^n=r^2 = 1 , r s r = s^{-1} \rangle$$ Our generators are already in this form, so set $s=f,g=r$}.

The action of the point group on bosons is given through \eqref{eq:spinproj} as
\begin{equation} \label{eq:bosd6}
    \begin{aligned}
        f_L &= \left(\begin{array}{ccc}
                 R(2 \pi/3) & 0 & 0 \\
                 0 & I_2 & 0 \\
                 0 & 0 & I_2 \\
                \end{array}\right) \,, \qquad
        &f_R =& \left(\begin{array}{ccc}
                 -I_2 & 0 & 0 \\
                 0 & -I_2 & 0 \\
                 0 & 0 & I_2 \\
                \end{array}\right) \, \\
        g_L &= \left(\begin{array}{ccc}
                \mathcal{P} & 0 & 0 \\
                 0 & -I_2 & 0 \\
                 0 & 0 & \mathcal{P} \\
                \end{array}\right) \,, \qquad
        &g_R =& \left(\begin{array}{ccc}
                -I_2 & 0 & 0 \\
                 0 & -I_2 & 0 \\
                 0 & 0 & I_2 \\
                \end{array}\right) \,. \\
    \end{aligned}
\end{equation}
with the same notation as in \eqref{eq:bosS3}. The lattice with this symmetry is the same as the previous Section, \eqref{eq:LatticeS3Z3}. The base circle allows us to mass up some twisted sectors by adding geometric shifts on it, with an extra subtlety compared to the previous Section. In this case the commutator subgroup of the point group is
\begin{equation}
    [D_6,D_6] \simeq \Z_3 \simeq \langle f^2 \rangle\,,
\end{equation}
so adding an order 6 shift on $f$ would lead to an enhancement of the space group, since on the base circle one would find $gfg^{-1}f^{-1} \neq f^2$. To solve this, we use an order 2 shift both on $f$ and $g$, which can be explicitly seen to preserve the group structure.

Then gravitini can only come from the commutator subgroup, for which the twist is
\begin{equation} \label{eq:commutator_d6}
    f^2: \qquad \vec{\theta}_L = \left(\frac{2}{3},0,0\right), \qquad \vec{\theta}_R = (0,0,0)\,.
\end{equation}
The situation is entirely analogous to that of the previous subsection, and it has the same solution: there would be a gravitino without adding further shifts, but since \eqref{eq:commutator_d6} is the inverse of \eqref{eq:commutator_twist} and $g$ acts the same way, we add to $f$ a shift which is the inverse of \eqref{eq:shiftS3}. The same calculation then shows that all twisted gravitini are massed up, and SUSY is truly broken.

To summarize, the bosonic action we choose is, for $f$:
\begin{equation}
    f:\left\{
    \begin{aligned}
        \left(
        \begin{array}{c}
             X_1^L \\
             X_2^L
        \end{array}
        \right)
        &\longmapsto
        \left(
        \begin{array}{cc}
         0 & -1 \\
         1 & -1 \\
        \end{array}
        \right)
        \left(
        \begin{array}{c}
             X_1^L \\
             X_2^L
        \end{array}
        \right) ,  & \quad
        \left(
        \begin{array}{c}
             X_1^R \\
             X_2^R
        \end{array}
        \right)
        &\longmapsto
        \left(
        \begin{array}{c}
             -X_1^R \\
             -X_2^R
        \end{array}
        \right)\,, \\ %end line here    
        \left(
        \begin{array}{c}
             X_3^L \\
             X_4^L
        \end{array}
        \right)
        &\longmapsto
        \left(
        \begin{array}{c}
             X_3^L \\
             X_4^L
        \end{array}
        \right)
        +\left(\begin{array}{c}
                1/3 \\
                2/3
              \end{array}\right), & \quad
        \left(
        \begin{array}{c}
             X_3^R \\
             X_4^R
        \end{array}
        \right)
        &\longmapsto
        \left(
        \begin{array}{c}
             -X_3^R \\
             -X_4^R
        \end{array}
        \right) \,,\\ %end line here    
        \left(
        \begin{array}{c}
             X_5^L \\
             X_6^L
        \end{array}
        \right)
        &\longmapsto
        \left(\begin{array}{c}
                X_5^L \\
                X_6^L
              \end{array}\right)
        +\left(\begin{array}{c}
                0 \\
                l_b/2
              \end{array}\right) ,  & \quad
        \left(
        \begin{array}{c}
             X_5^R \\
             X_6^R
        \end{array}
        \right)
        &\longmapsto
        \left(\begin{array}{c}
                X_5^R \\
                X_6^R
              \end{array}\right)
        +\left(\begin{array}{c}
                0 \\
                l_b/2
              \end{array}\right),
    \end{aligned}
    \right.
\end{equation}
while for $g$ we have the same as in \eqref{eq:gS3boson}.

As a last comment on non-Abelian point groups, we point out that the fermionic action \eqref{eq:fd6} for $f$, with $\zeta$ a root of unity of any order, $\zeta^n=1$, can be paired with $g$ as in \eqref{eq:nonAbPoint} to give a fermionic rep which satisfies conditions \textbf{I.-IV.} of \hyperref[sec:groupth]{Section 3}. This gives a faithful representation of a dihedral group $D_n$ of arbitrarily high order which would realize local SUSY conservation. We emphasize however that the action  is non-geometric, so this cannot be used to give a large $n$ family of non-compact, local orbifold singularities with SUSY preserved to one-loop, because the splitting $X=X_L+X_R$ which is essential for the asymmetric orbifold construction does not make sense if $X$ is non-compact. Even with a compact internal manifold, we are restricted by condition \textbf{V.}, since the crystallographic classification of lattice automorphisms forbids finding a lattice with symmetries that lift to $f$ and $g$ for arbitrary $n$. This example clearly shows why a full classification of non-Abelian groups that satisfy our conditions is more involved, since one cannot put simple bounds on e.g. the order of the group itself. We thus leave a full classification of such groups in the context of toroidal orbifolds to future work.

\section{Global anomalies for non-Abelian models}\label{sec:anomalies}
In this section we give some details on the non-Abelian anomalies discussed in \hyperref[sec:asymmOrb]{Section 2} and \hyperref[sec:models]{Section 5}, explaining why they are the only ones that can possibly appear. Orbifolding a symmetry means gauging it in the worldsheet; in a type II model, and absent RR fields, branes and orientifolds, a consistent background is specified by a $(1,1)$ 2d SCFT with an anomaly-free internal $\mathbb{Z}_2$ symmetry, necessary for the GSO projection. If we wish to gauge some global symmetry $G$ to construct an orbifold model, there is an anomaly for $G$, which is captured by a three-dimensional bordism group $\Omega_3^{\text{Spin}}(B\mathbb{Z}_2\times BG)$ in case the internal symmetry $G$ does not include $(-1)^F$ as a subgroup. We will now study these bordism groups and anomalies for the cases of interest.

\subsection{Non-Abelian anomaly for $\Delta_{27}$}

Global anomalies of fermions living in $d$ dimensions charged under a gauge group $G$ are bordism invariants(see e.g. \cite{Garcia-Etxebarria:2018ajm,Davighi:2022icj}):
\begin{equation}
    \Hom(\Omega_{d+1}^\Spin(BG),U(1)),
\end{equation}
So to classify potential global anomalies we have to compute the relevant bordism groups.
For $G = \Delta_{27}$, we will follow closely the approach of \cite{Freed:1987qk}, which discussed the $\mathbb{Z}_5$ case. We can use the Atiyah-Hirzebruch spectral sequence (AHSS) to express the result in terms of group cohomology, which is known in the literature \cite{Leary_1991}. Using the fibration $pt \rightarrow B\Delta_{27} \rightarrow B\Delta_{27}$,  the $E^2$ page of the AHSS reads
\begin{equation}
    E^2_{p,q} = H_p(B\Delta_{27}; \Omega_{q}^\Spin(pt))\,.
\end{equation}
The bordism groups of a point read \cite{Anderson-Brown-Peterson}
\begin{table}[h]
    \centering
    \begin{tabular}{c|c|c|c|c|c}
        
        $d$ & 0 & 1 & 2 & 3 & 4 \\
        \hline
        $\Omega_d^{\text{Spin}}(pt)$ & $\mathbb{Z}$ & $\mathbb{Z}_2$ & $\mathbb{Z}_2$ & 0 & $\mathbb{Z}$\\
       
    \end{tabular}
    \label{tab:spin_bordism}
\end{table}
\\Furthermore, $\Delta_{27}$ is a finite group of odd order, so its homology only has components that are torsion of odd order. This, together with the universal coefficient theorem (see e.g. \cite{MR1867354}), is enough to write
\begin{equation}
    H_p(B\Delta_{27};\Z_2) = 0\,,
\end{equation}
so the second page of the AHSS reads, in low degrees:
\begin{equation}
    E^2_{p,q} : \quad
    \begin{array}{c|cccc}
        3 & & & & \\
        2 & \mathbb{Z}_2 & & & \\
        1 & \mathbb{Z}_2 & & & \\
        0 & \mathbb{Z} & H_1(B\Delta_{27}) & H_2(B\Delta_{27}) & H_3(B\Delta_{27}) \\
        \hline
         p/q & 0 & 1 & 2 & 3
    \end{array}
\end{equation}
From it we can see that in total degree 3 there is only one contribution $E_{3,0} \simeq H_3(B\Delta_{27})$. All differentials in these degrees vanish trivially, so we find
\begin{equation} \label{eq:bordDelta}
    \Omega_3^\Spin(B\Delta_{27})\simeq H_3(B\Delta_{27}) \simeq H^4(\Delta_{27};\Z) \simeq \Z_3 \oplus \Z_3 \oplus \Z_3 \oplus \Z_3\,,
\end{equation}
where the second isomorphism is given by the universal coefficient theorem again, and the group cohomology was calculated for the generic Heisenberg group over $\Z_p$ in \cite{Leary_1991}.
This exact same procedure can be applied to its maximal Abelian subgroup $ P_G =\Z_3 \times \Z_3$, giving
\begin{equation}\label{eq:bordZ3Z3}
    \Omega_3^\Spin(B\Z_3 \times B \Z_3)\simeq \Z_3 \oplus \Z_3 \oplus \Z_3 \,.
\end{equation}
These three factors are straightforwardly interpreted as the Abelian anomalies associated to cyclic subgroups inside $\Z_3 \times \Z_3$, which are cancelled in level matched models \cite{Freed:1987qk}.

The discrepancy between \eqref{eq:bordDelta} and \eqref{eq:bordZ3Z3} signals that there is an extra potential anomaly in a $\Delta_{27}$-theory which is not detected by restricting to maximal Abelian subgroups. Following \cite{Freed:1987qk}, the extra anomalies in non-Abelian groups are given by the difference in phase between sectors $Z[f,g]$ and $Z[f^ag^b,f^cg^d]$ for which there exists an element $h\in G$ such that
\begin{equation}
    (hfh^{-1},hgh^{-1}) = (f^ag^b,f^cg^d)
\end{equation}
For $\Delta_{27}$, this can happen in the unique faithful irreducible representation, which did show up in the models of \hyperref[sec:models]{Section 5}. This representation is generated by
\begin{equation}
    f = \left(
\begin{array}{ccc}
0 & 1 & 0 \\
0 & 0 & 1 \\
1 & 0 & 0
\end{array}
\right), \quad g = \left(
\begin{array}{ccc}
\phi & 0 & 0 \\
0 & \phi^2 & 0 \\
0 & 0 & 1
\end{array}
\right), \quad c = f g f^{-1} g^{-1} =\left(
\begin{array}{ccc}
\phi & 0 & 0 \\
0 & \phi & 0 \\
0 & 0 & \phi
\end{array}
\right) 
\end{equation}
where $\phi^3=1$ is a third root of unity. The physical meaning of the anomaly is as follows: $f$ conjugates $(f,g)$ to $(f,fg)$, so if the symmetry is gauged, the two should have the same phase for the partition function. However, the formulas in Section 3 of \cite{Freed:1987qk} give a phase $e^{2 \pi i/3}$ between this two sectors, signaling that a single copy of this representation would break modular invariance. In fact, the analogous case for the faithful representation of the Heisenberg group over $\Z_5$ is performed in \cite{Freed:1987qk}, and finds an analogous anomaly.
It is therefore fundamental for us to have this irreducible representation show up in the same way on the left and on the right, so that the full representation is effectively non-chiral.

\subsection{Non-Abelian point groups}

For the non-Abelian point group models, from the get go we could in principle have additional anomalies not fixed by level matching. We show by computing the relevant bordism groups that this is not the case, and so level matching is enough to guarantee modular invariance.

Let us start from $G = S_3 \times \Z_3$. The bordisms of individual factors are known \cite{Davighi:2022icj}:
\begin{equation}
    \begin{aligned}
        \Omega^\text{Spin}_3(B \Z_3) &\simeq \Z_3 \\
        \Omega^\text{Spin}_3(B S_3) &\simeq \Z_3 \oplus \Z_8 \simeq\Omega^\text{Spin}_3(B \Z_3 \times B\Z_2)
    \end{aligned}
\end{equation}
The second of these has the interpretation that $S_3$ global anomalies are only those inherited by the cyclic subgroups $\Z_3$ and $\Z_2$.
When taking the direct product $S_3 \times \Z_3$, there could in principle be more components arising in the bordism group, but from \cite{Braeger:2025kra} we see that
\begin{equation}
    \Omega^\Spin_3(BS_3\times B \Z_3) \simeq \Omega^\text{Spin}_3(B S_3) \oplus \Omega^\text{Spin}_3(B \Z_3)
\end{equation}
when localized at prime 3, and the splitting at prime 2 is trivial since $\Omega_3^\Spin(B\Z_3)$ has no order 2 component. All together, this tells us that there are no extra non-Abelian anomalies, so level matched $S_3 \times \Z_3$ models are modular invariant.

For $G= D_6$, the calculation is more involved. It is convenient to use the isomorphism $D_6\simeq S_3 \times \Z_2$. From this, at prime 3 we see that
\begin{equation}
    \Omega^\Spin_3(BD_6)\big{|}_{p=3} \simeq \Omega^\Spin_3(BS_3)\big{|}_{p=3} \simeq \Z_3,
\end{equation}
since $\Z_3$-valued homology of $B\Z_2$ vanishes.
For the prime 2 factors, we can again use the AHSS. Homology of the classifying spaces in low degrees can be computed through Künneth's theorem \cite{MR1867354}:
\begin{equation}
    H_p(BD_6;\Z) \simeq
    \begin{cases}
        \Z \quad &p = 0 \\
        \Z_2^{\oplus 2} \quad &p= 1,4\\
        \Z_2 \oplus \Z_6 \quad  &p = 2 \\
        \Z_2^{\oplus 3} \quad &p=3
    \end{cases} \,,
    \qquad \quad
    H_p(BD_6;\Z_2) \simeq \Z_2^{\oplus(p+1)}
\end{equation}
From this we can build the second page of the AHSS, which ignoring order 3 factors reads:
\begin{equation}
    E^2_{p,q} : \quad
    \begin{array}{c|ccccc}
        3 & & & & &\\
        2 & \, \, \mathbb{Z}_2 \, \, & \, \, \mathbb{Z}_2^{\oplus 2} \, \,& \, \, \mathbb{Z}_2^{\oplus 3} \, \, & \, \, \mathbb{Z}_2^{\oplus 2} \, \,& \, \, \Z_2^{\oplus 3} \, \,\\
        1 & \, \, \mathbb{Z}_2 \, \, & \, \, \mathbb{Z}_2^{\oplus 2}  \, \,& \, \, \mathbb{Z}_2^{\oplus 3} \, \,  & \, \, \mathbb{Z}_2^{\oplus 2} \, \,& \, \, \Z_2^{\oplus 2} \, \,\\
        0 & \, \, \mathbb{Z} & \mathbb{Z}_2^{\oplus 2} \, & \mathbb{Z}_2 & \mathbb{Z}_2^{\oplus 3} \, & \, \, \Z_2^{\oplus 2} \, \,\\
        \hline
         p/q & 0 & 1 & 2 & 3 & 4
    \end{array}
\end{equation}
From this spectral sequence, the order of the 2-torsion in $\Omega^{\text{Spin}}_3(BD_6)$ is at most 256.

%use that the restriction map induced by inclusion $\Z_2 \times \Z_2 \rightarrow S_3 \times \Z_2$ is an isomorphism in group cohomology $H^(S_3\times\Z_2)\xrightarrow{\sim} H^*(\Z_2 \times \Z_2)$ since $\Z_2 \times \Z_2$ is a Sylow 2-subgroup \cite{brown1982cohomology}.

The inclusion map 
\begin{equation}\mathbb{Z}_2\times\mathbb{Z}_2\,\rightarrow\, S_3\times \mathbb{Z}_2,\end{equation}
which sends the generator of the first $\mathbb{Z}_2$ to any order 2 permutation,\footnote{It does not matter which one chooses as all Sylow 2-subgroups are conjugate to each other \cite{brown1982cohomology}. Because of this, there is an isomorphism in group cohomology $H^*(S_3\times\Z_2)\xrightarrow{\sim} H^*(\Z_2 \times \Z_2)$.}
induces a homomorphism of the corresponding spin bordism groups. In fact, it is an isomorphism, since all classes in the bordism group \cite{Guo:2018vij} 
\begin{equation} \label{eq:2}
    \Omega^\Spin_3( B\Z_2 \times B\Z_2) \simeq \Z_8^{\oplus 2} \oplus \Z_4 
\end{equation}
are detected by $\eta$ invariants corresponding to $\mathbb{Z}_2\times \mathbb{Z}_2$ representations, all of which survive as $S_3\times \mathbb{Z}_2$ representations. As a result, all classes in \eq{eq:2} survive as classes in $\Omega^{\text{Spin}}_3(BD_6)$, and since the order of this group is 256, the bound is saturated. We conclude that the full bordism group reads
\begin{equation}
    \Omega^\Spin_3(BD_6) \simeq \Z_3 \oplus \Z_8^{\oplus 2} \oplus \Z_4
\end{equation}
These are all anomalies of Abelian subgroups. Indeed, the result obtained coincides with the bordism group
\begin{equation} 
    \Omega^\Spin_3(B\Z_3 \times B\Z_2 \times B\Z_2) \simeq \Z_3 \oplus \Z_8^{\oplus 2} \oplus \Z_4 \,,
\end{equation}
This means that level matching is enough to guarantee modular invariance also for $D_6$ models.

\section{So, why did we end up working on these asymmetric orbifolds?}\label{sec:conclus}

In \cite{Heckman:2024obe,Kaidi:2024wio}, it was shown how non-invertible symmetries in perturbative string theory can yield selection rules which are valid at tree level, but which are generically broken at higher loops. This discussion however did not apply to the vacuum energy, since it was all about internal symmetries.  On the other hand, the mechanism proposed in \cite{Kachru:1998hd} to ensure vanishing vacuum energy at one loop shares many superficial similarities with the broken non-invertible symmetries studied in \cite{Heckman:2024obe, Kaidi:2024wio}; we started to study it to see if we could find a mechanism analogous to  the one discussed in this reference, possibly involving something like a ``non-invertible supersymmetry'', to construct models with vanishing vacuum energy at $g$ loops. This would be very interesting as a mechanism to engineer anomalously small vacuum energies without supersymmetry.

However, we got sidetracked early on. As noted already in \cite{Harvey:1998rc} and explained in detail in \cite{Aoki:2003sy}, the argument of \cite{Kachru:1998hd} had some subtle issues, since for an asymmetric orbifold the equivalence with a simpler, non-compact $\R^n/\Tilde{G}$ model is not in general possible. Although the model does have a vanishing one-loop vacuum energy, we were left wondering whether one could construct models realizing the principle introduced in \cite{Kachru:1998hd}, which is so similar to the way internal non-invertible symmetries are broken by loops. The models in the present paper fill in this gap: their vanishing 1-loop vacuum energy can be understood as arising from supercharges that are defined in each twisted sector, but not globally, and the theory we have is a finite order orbifold of a toroidal model (see \cite{Satoh:2015nlc} for a similar mechanism using infinitely many winding sectors, and \cite{Aoyama:2020aaw} for a construction involving Gepner models where the vacuum energy is tuned to vanish). 

We have classified all possible Abelian point groups which can appear in models with a vanishing one-loop vacuum energy. It would be interesting to extend this classification to non-Abelian point groups, for which we only studied a couple of concrete examples, or to more general (non-toroidal) orbifolds. Both possibilities seem considerably harder.

An outstanding question is whether the vacuum energy vanishes beyond one-loop in our models. Reference \cite{Aoki:2003sy}, which found a non-vanishing 2-loop vacuum energy for the model of \cite{Kachru:1998hd}, attributed this to a failure to realize the structure of supercharge-like operators sector by sector, while still breaking spacetime SUSY. Since our models realize this, at least in the worldsheet fermionic sector responsible for spacetime supersymmetry, it is natural to wonder whether the cancellation could survive at higher loops. Of course, a detailed calculation is needed to establish this. Related to this, it is important to emphasize that, in string perturbation theory, the first non-vanishing contribution to the vacuum energy is readily computable, but since it depends on some power of $g_s$, it generates a dilaton tadpole and the higher-order corrections must be computed around a shifted vacuum, following the recipe in \cite{Fischler:1986ci} (see also \cite{Polchinski_1998} for a pedagogical account).

In any case, in all of these models, a low-energy effective field theorist would find a bizarre suppression of the vacuum energy, due to a non-supersymmetric spectrum that looks fine tuned,  which would be very difficult to explain from the EFT point of view -- yet it is completely natural to a string theorist with access to the worldsheet CFT. In this sense, these models can be viewed as a counterpart of \cite{Angelantonj:2003hr, Dudas:2025yqm}, in which the surprising cancellations take place in the closed-string sector. Understanding this and similar phenomena might hold the key to understanding the smallness of the vacuum energy in our world.

\section*{Acknowledgments}
 We thank Carlo Angelantonj, Ivano Basile, Zihni Kaan Baykara, Ralph Blumenhagen, Bernardo Fraiman, Salvatore Raucci, Yuji Satoh, Houri-Christina Tarazi, and Cumrun Vafa for useful discussions, correspondence, and comments. We are also grateful to Ethan Torres and Irene Valenzuela for discussions and ongoing collaboration on a project from which this one branched out. 
 MT is supported by the FPI grant PRE2022-102286 from Spanish National Research Agency from the Ministry of Science and Innovation. MM and MT thank the Spanish Research Agency
(Agencia Estatal de Investigacion) through the grants IFT Centro de Excelencia Severo
Ochoa CEX2020-001007-S, PID2021-123017NB-I00 and PID2024-156043NB-I00, funded by
MCIN/AEI/10.13039/501100011033 and by ERDF A way of making Europe. MM is currently supported by the RyC grant RYC2022-037545-I and project EUR2024-153547 from the AEI. VL thanks the IFT for the hospitality during the earlier steps of this work.

\appendix

\section{Symmetries of lattices}\label{app:A}

Symmetries of lattices are very constrained in terms of their eigenvalues. These can be studied by using cyclotomic polynomials. See \cite{Baykara:2024vss,Baykara:2025gcc} for recent discussions in the physics literature. We summarize the result here briefly, for more details see \cite{book:93106273,Kuzmanovich01022002}. Consider a rank $n$ lattice $\Lambda$ and an automorphism $g \in \text{Aut}(\Lambda)$. Choosing a lattice basis, this is represented by a $n \times n$ integral matrix $G \in GL(n;\Z)$. Its eigenvalues are found as roots of the minimal polynomial of this matrix, call it $\mu_G$. One of its basic properties is that it divides all polynomials $P$ that annihilate $G$, i.e. those for which $P(G)=0$.
\\For an order $k$ automorphism $g$, its matrix is  annihilated by $P(x)=x^k-1$.

The $k$-th cyclotomic polynomial $\Phi_k(x)$ is precisely defined as the unique irreducible polynomial with integer coefficients which divides $x^k-1$ but does not divide $x^{k'}-1$ for any $k'<k$. This has the consequence that 
\begin{equation}
    x^k - 1 = \prod_{d|k} \Phi_d(x)
\end{equation}
For prime $k$ this is particularly easy, and reads
\begin{equation}
    x^k-1 = (x-1) \Phi_k(x)
\end{equation}
Since both factors are irreducible, one of them has to be $\mu_G$. This means that the eigenvalues of an order $k$ automorphism are either all $1$, or they are found as the roots of the cyclotomic polynomial $\Phi_k$.

The first few cyclotomic polynomials read as follows:
\begin{equation}   \label{eq:cyclotomic}
    \begin{aligned}
        \Phi_1(x) &= x - 1 \\
        \Phi_2(x) &= x + 1 \\
        \Phi_3(x) &= x^2 + x + 1 \\
        \Phi_4(x) &= x^2 + 1 \\
        \Phi_5(x) &= x^4 + x^3 + x^2 + x + 1 \\
        \Phi_6(x) &= x^2 - x + 1 \\
        \Phi_7(x) &= x^6 + x^5 + x^4 + x^3 + x^2 + x + 1 \\
        \Phi_8(x) &= x^4 + 1 \\
        \Phi_9(x) &= x^6 + x^3 + 1 \\
        \Phi_{10}(x) &= x^4 - x^3 + x^2 - x + 1
    \end{aligned}
\end{equation}
From these one can read off our cases of interest. In particular, for $\Z_5$ actions we see that the symmetry must act either trivially or on a 4-dimensional block, with eigenvalues
\begin{equation}
    (\omega,\omega^2,\omega^3,\omega^4) \qquad \text{for} \qquad \omega = e^{2 \pi i/5} \,,
\end{equation}
and with the same reasoning a $\Z_7$ symmetry acts either trivially or on a 6-dimensional block, with eigenvalues
\begin{equation}
    (\xi,\xi^2,\xi^3,\xi^4,\xi^5,\xi^6) \qquad \text{for} \qquad \xi = e^{2 \pi i/7} \,.
\end{equation}

\section{Toroidal partition functions and Theta functions}\label{app:TorusZ}

In this Appendix we want to prove that the 1-loop partition function of a toroidal orbifold vanishes in each sector where a gravitino-like operator is preserved, and only when that is the case.
In a toroidal orbifold, the 1-loop partition function can be written as a sum over sectors as in \eqref{eq:orbifold_partition_function}. For a type II compactification on $T^6$, the partition function factorizes into left and right movers, and each of the two factors has the form:
\begin{equation}
    \mathcal{Z}[f,g]= \mathcal{Z}_{\mathbb{R}^{1,3}} \cdot \mathcal{Z}^B_{T^6}[f,g] \cdot \mathcal{Z}^F[f,g]
\end{equation}
where $\mathcal{Z}_{\mathbb{R}^{1,3}}$ is the universal factor associated to the external bosons, $\mathcal{Z}^B_{T^6}[f,g]$ is the partition function of the compact bosons, and
\begin{equation} \label{eq:zFermions}
    \mathcal{Z}^F[f,g] = \mathcal{Z}^{NS}[f,g] - \mathcal{Z}^R[f,g]
\end{equation}
is the partition function of the worldsheet fermions.
This last factor is the only one that can vanish, since the rest are partition functions of unitary bosonic CFT's, and thus are strictly positive on their own.

Writing the twist vector of $f$ as $(f_1,f_2,f_3)$ and similarly for $g$, the two summands read (see e.g.\cite{Gkountoumis:2023fym}):
\begin{equation}\label{eq:twisted_partition_functions}
    \begin{aligned}
        \mathcal{Z}_{NS}[f,g] &= \frac{1}{2}e^{\pi i \sum_j f_j g_j} \left[ \left( \frac{\vartheta_3}{\eta} \right) \prod_j \frac{\vartheta \left[ f_j \atop -g_j \right]}{\eta} -e^{\pi i \sum_j f_j} \left( \frac{\vartheta_4}{\eta} \right) \prod_j \frac{\vartheta \left[ f_j \atop -\frac{1}{2} - g_j \right]}{\eta}     \right] \, ,\\
        \mathcal{Z}_{R}[f,g] &= \frac{1}{2}e^{\pi i \sum_j f_j g_j} \left[ \left( \frac{\vartheta_2}{\eta} \right) \prod_j \frac{\vartheta \left[ \frac{1}{2}+f_j \atop -g_j \right]}{\eta} + e^{\pi i \sum_j f_j} \left( \frac{\vartheta_1}{\eta} \right) \prod_j \frac{\vartheta \left[ \frac{1}{2} + f_j \atop  - \frac{1}{2} -g_j \right]}{\eta}     \right] \,.
    \end{aligned}
\end{equation}
The theta functions with characteristics are defined as
\begin{equation}\label{eq:thetadef}
        \vartheta \left[\alpha \atop \beta \right] (z | \tau) = \sum_{n \in \Z} q ^ \frac{(n+\alpha)^2}{2} e^{2 \pi i (n+\alpha)(z+\beta)}, \qquad \text{with } \, q = e^{2 \pi i \tau}
\end{equation}
For ease of notation, we will always omit the $\tau$ argument, and when the $z$ argument is omitted as well, it is implicitly set to zero. The special values $\alpha,\beta = 0, 1/2$ define the four classical theta functions:
\begin{equation}\label{eq:theta_pro1}
    \begin{aligned}
        \vartheta_3(z) &:= \vartheta \left[0 \atop 0 \right] ( z | \tau),& \vartheta_4(z) &:= \vartheta \left[0 \atop 1/2 \right] ( z | \tau), \\ \vartheta_2(z) &:= \vartheta \left[1/2 \atop 0 \right] ( z | \tau), & \vartheta_1(z) &:= \vartheta \left[1/2 \atop 1/2 \right] ( z | \tau)
    \end{aligned}
\end{equation}
Some useful basic properties of the $\vartheta$-functions  are
\begin{equation}\label{eq:theta_prop2}
    \vartheta \left[-\alpha \atop -\beta \right] (0 | \tau) = \vartheta \left[\alpha \atop \beta \right] (0 | \tau), \quad \vartheta \left[\alpha + n \atop \beta +m \right] (0 | \tau) = e^{2 \pi i m \alpha}\vartheta \left[\alpha \atop \beta \right] (0 | \tau)
\end{equation}

We are going to focus here on the untwisted sector by setting $f_j=0$, and we factor out common non-zero prefactors in $\mathcal{Z}^F(\tau,\bar{\tau})$ since we are only interested in whether the partition function vanishes or not. The quantity we wish to compute is then
\begin{equation}\label{eq:thetaCombo}
    \vartheta_3 \prod_j \vartheta \left[ \text{\small{$0$}} \atop \text{\small{$-g_j$}} \right] - \vartheta_4 \prod_j \vartheta \left[ \text{\small{0}} \atop \text{\small{$-1/2-g_j$}} \right] - \vartheta_2 \prod_j \vartheta \left[ \text{\small{$1/2$}} \atop \text{\small{$\text{\small{$-g_j$}}$}} \right] - \vartheta_1 \prod_j \vartheta \left[ \text{\small{$1/2$}} \atop \text{\small{$-1/2-g_j$}} \right]
\end{equation}
The last summand contains a $\vartheta_1 = 0$ factor, but for now we keep it explicitly to compare this with known formulas.
\\We make use of the following identities, which can be directly checked with \eqref{eq:thetadef}:
\begin{equation}\label{eq:thetaIdentities}
    \begin{aligned}
        \vartheta \left[ 0 \atop \beta \right](0) = \vartheta_3(\beta) = \vartheta_4(\beta+1/2), \\
        \vartheta \left[ 1/2 \atop \beta \right](0) = \vartheta_2(\beta) = \vartheta_1(\beta+1/2)
    \end{aligned}
\end{equation}
to write \eqref{eq:thetaCombo} as
\begin{equation}\label{eq:thetaZ}
    \vartheta_3(0) \prod_j \vartheta_3(-g_j) - \vartheta_4(0) \prod_j \vartheta_4(-g_j) - \vartheta_2(0) \prod_j \vartheta_2(-g_j) - \vartheta_1(0) \prod_j \vartheta_1(-g_j)  
\end{equation}
There are analogous relations to \eqref{eq:thetaIdentities} that can be used in the twisted sector: 
\begin{equation}\label{eq:thetaIdentitiesII}
    \begin{aligned}
        &\vartheta \left[ \alpha \atop 0 \right](0) = e^{i\pi \tau \alpha^2} \vartheta_3(\alpha\,\tau) = e^{i\pi\tau\left(\alpha-\frac{1}{2}\right)^2}\vartheta_2\left(\left(\alpha-\tfrac{1}{2}\right)\,\tau\right),\\
        &\vartheta \left[ \alpha \atop 1/2 \right](0) = e^{i\pi \tau \alpha^2}e^{i\pi\alpha}\vartheta_4(\alpha\,\tau) = e^{i\pi \tau \left(\alpha-\frac{1}{2}\right)^2}e^{i\pi\left(\alpha-\frac{1}{2}\right)}\vartheta_1(\left(\alpha-\tfrac{1}{2}\right)\,\tau) \,.
    \end{aligned}
\end{equation}
There are many identities relating various quartic expressions in the $\vartheta$'s. The one we use here is (see equation (3.17) of \cite{KHARCHEV201519}):
\begin{equation} \label{eq:RiemannQuartic}
    \begin{aligned}
    \vartheta_1(a)\vartheta_1(b)\vartheta_1(c)\vartheta_1(d) \,+& \,\vartheta_2(a)\vartheta_2(b)\vartheta_2(c)\vartheta_2(d)\, - \\
    - \, \vartheta_3(a)\vartheta_3(b)\vartheta_3(c)\vartheta_3(d) \, + \, & \, \vartheta_4(a)\vartheta_4(b)\vartheta_4(c)\vartheta_4(d) = 2 \, \vartheta_1(a')\vartheta_1(b')\vartheta_1(c')\vartheta_1(d') \, ,
    \end{aligned}
\end{equation}
where the primed coordinates are obtained as the linear combinations
\begin{equation}\label{eq:theta&spin}
    \left(
        \begin{array}{c}
         a' \\
         b' \\
         c' \\
         d' \\
        \end{array}
    \right) = 
    \frac{1}{2}\left(
        \begin{array}{cccc}
         -1 & 1 & 1 & 1 \\
         1 & -1 & 1 & 1 \\
         1 & 1 & -1 & 1 \\
         1 & 1 & 1 & -1 \\
        \end{array}
    \right)
    \left(
        \begin{array}{c}
         a \\
         b \\
         c \\
         d \\
        \end{array}
    \right) \,.
\end{equation}
Note that  setting $(a,b,c,d)=0$, \eqref{eq:RiemannQuartic} becomes the famous Jacobi abstruse identity:
\begin{equation}
    \vartheta_2^4 - \vartheta_3^4 + \vartheta_4^4 = 0 \,.
\end{equation}

The matrix \eqref{eq:theta&spin} is exactly the same linear relation between the entry of the twist vector $g_i$ of an $SO(8)$ transformation on the torus coordinates $X^\mu$, and their spin lift which expresses the (phases) of the action on the supercharges, as explained around \eqref{eq:spinlift} in the case of $SO(6)$. On a $T^6$, we then use \eqref{eq:RiemannQuartic} with $a=0, (b,c,d) = -(g_1,g_2,g_3)$ to express \eqref{eq:thetaCombo} as a product of $\vartheta_1$'s.

Then by the basic property
\begin{equation}\label{eq:theta_prop3}
    \vartheta_1(n|\tau) = 0 \qquad \text{for } n \in \Z \, ,
\end{equation}
we see that the partition function vanishes whenever the twists are such that at least one eigenvalue $(a',b,c',d')$ in the spin lift is trivial (i.e. integral), or equivalently whenever a supercharge is conserved in that sector. Thus,  the complicated identity \eqref{eq:thetaCombo} may be regarded as a consequence of supersymmetry!

For the non-Abelian example in \hyperref[sec:d6]{Section 5.4}, to show the vanishing of the partition function we need to compute an extra sector which is not directly obtained through modular transformations from the untwisted sector. In particular, we need to compute an operator insertion in the sector with boundary conditions twisted by spacetime fermion number $(-1)^F$. This will change the sign in front of the Ramond sector in \eqref{eq:zFermions} to give
\begin{equation}
    \mathcal{Z}^F[(-1)^F,g] = \mathcal{Z}^{NS}[1,g]+\mathcal{Z}^{R}[1,g]
\end{equation}
To see if this vanishes, we need to compute the equivalent of \eqref{eq:thetaZ} in this sector, which reads:
\begin{equation}\label{eq:thetaZ2}
    \vartheta_3(0) \prod_j \vartheta_3(-g_j) - \vartheta_4(0) \prod_j \vartheta_4(-g_j) + \vartheta_2(0) \prod_j \vartheta_2(-g_j) + \vartheta_1(0) \prod_j \vartheta_1(-g_j)  
\end{equation}
To see when this vanishes, we make use of another identity from \cite{KHARCHEV201519}:
\begin{equation} \label{eq:RiemannQuartic2}
    \begin{aligned}
    \vartheta_1(a)\vartheta_1(b)\vartheta_1(c)\vartheta_1(d) \,+& \,\vartheta_2(a)\vartheta_2(b)\vartheta_2(c)\vartheta_2(d)\,+ \\
    +\,\vartheta_3(a)\vartheta_3(b)\vartheta_3(c)\vartheta_3(d)\,- \, & \, \vartheta_4(a)\vartheta_4(b)\vartheta_4(c)\vartheta_4(d) = 2 \, \vartheta_2(a')\vartheta_2(b')\vartheta_2(c')\vartheta_2(d') \, .
    \end{aligned}
\end{equation}
Now we have $\vartheta_2$'s on the right hand side, for which
\begin{equation}
    \vartheta_2(n+\tfrac{1}{2} \, | \, \tau) = 0 \qquad \text{for } n \in \Z
\end{equation}
holds. Then by recalling that the primed entries are nothing but the phases of the eigenvalues of the fermionic action of the twist $g$ we obtain that the partition function $\mathcal{Z}^F[(-1)^F,g]$ vanishes whenever $g$ has an eigenvalue $-1$. This agrees with the intuitive expectation that fermions in the $(-1)^F$-twisted sector somehow behave with the opposite eigenvalues.

\section{Reabsorbing shifts} \label{app:shifts}
The constructions presented in the main text rely on the inclusion of shifts with components not in the invariant lattice $I$, which are usually not present in the literature. In this Appendix we briefly discuss why this is the case. The point is that through a basic redefinition one can reabsorb the components of a shift which are orthogonal to $I$. Take an affine transformation
\begin{equation}
    X \mapsto X' = \theta X+v.
\end{equation}
By setting $Y = X+b$ one finds
\begin{equation}
    Y' = X' + b = \theta (Y-b) + v + b = \theta Y + v - (\theta-1)b.
\end{equation}
In order to reabsorb the shift and get a pure rotation one needs to solve for $b$ in 
\begin{equation}\label{eq:absorbshift}
    (\theta-1)b = v\,.
\end{equation}
Decompose $ v = v^* + v^\perp $ into its invariant component $ v^* \in I $ and its orthogonal complement, and the same for $b = b^* + b^\perp$. It is then clear that we can solve \eqref{eq:absorbshift} only in the $I^\perp$ components, and uniquely so given $\theta$ and $v$.

A large fraction of the literature focuses on cyclic orbifolds. For these, one can reabsorb a shift in the non-invariant components by \eqref{eq:absorbshift}, which is why shifts are taken directly in the invariant lattice $I$. Since we use groups with two generators, by uniqueness of the solution of \eqref{eq:absorbshift} it is impossible to reabsorb shifts for both generators if their twists $\theta$ are different. This means that we can without loss of generality set the shift in $I^\perp$ on one generator to be trivial, but the other can be chosen as to make the full group genuinely non-Abelian on bosons.

\section{Commuting pairs in non-Abelian orbifolds} \label{app:D}

In this Appendix we verify that there are no accidental non-vanishing contributions to the partition function of our orbifolds, due to commuting elements which are not immediately seen by our presentation in terms of generators $f,g$.

For the example in \hyperref[sec:z3example]{Section 5.1} we argued that the full space group is given by the Heisenberg group over the field $\Z_3$. For the purpose of writing general commutators, it is convenient to use the isomorphism spelled out in \hyperref[footnote1]{Footnote 7} and use the generators $x,y,z$. $z$ is central, so one can write all elements as ordered products $x^ay^b z^n$. From this the general commutation relation is given as
\begin{equation}
    \left[x^ay^b z^n, x^c y^d z^m  \right] = z^{ad-bc}\,.
\end{equation}
Since all generators are order 3, two elements commute if $ad-bc = 0$ mod 3, or equivalently if $(a,b)$ and $(c,d)$ are linearly dependent if seen as vectors in $\Z_3\oplus \Z_3$. This means that all commuting pairs are of the form $h^a z^n, h^b z^m$ for some element $h$ in the group. These are extra sectors that we need to include in the orbifold sum. The non-trivial action of the translation $z$ on bosons makes it such that they are not in the same modular orbit as $\mathcal{Z}[1,h]$, so we need to add them by hand in the sum.
On fermions, however, shifts act trivially, so that
\begin{equation}
    \mathcal{Z}^F[h^a z^n, h^b z^m] = \mathcal{Z}^F[h^a,h^b]\,,
\end{equation}
which is indeed in the same modular orbit of $\mathcal{Z}^F[1,h]$, so it vanishes by construction. Thus all extra terms in the sum coming from ``accidental" commuting pairs vanish as well.

For the example in \hyperref[sec:z4example]{Section 5.2} it is again convenient to use the presentation as $(\Z_4 \times \Z_4) \rtimes \Z_4$, with the notation of \hyperref[fn:Z2xZ2xZ2xZ4]{Footnote 8}. The non-trivial commuting pairs certainly contain of any combination of the two generators $s_1,s_2$ in the $\Z_4 \times \Z_4$ factors. Furthermore, $z = s_1^2 s_2^2$ is central, and order 2 rotations $r^2$ and $r^2 s_1^n s_2^m$ commute with any order 2 shift $s_1^{2a} s_2^{2b}$.

We need to examine the action on fermions case by case.
First, $s_1$ and $s_2$ only different by shifts on bosons, so the action on fermions is the same as $g$. Thus in the $\Z_4 \times \Z_4$ factor we compute
\begin{equation}
    \mathcal{Z}^F[s_1^a s_2^b, s_1 ^c s_2 ^d] = \mathcal{Z}^F[g^{a+b},g^{c+d}] = 0\,,
\end{equation}
because it is in the modular orbit of $\mathcal{Z}^F[1,g]$ which vanishes by construction.
The central element $z$ acts by pure shift on bosons, so it does not act on fermions. Thus any sector that vanished before $z$ insertions will still vanish afterwards.
With the order 2 elements the situation is subtler. We compute the twist vectors for $r^2 = f^2, s_1^2= g^2, s_2^2 = fg^2f^{-1}$:
\begin{equation}
    \begin{aligned}
        f^2 : \qquad \vec{\theta}_L = \left(0,\frac{1}{2},\frac{1}{2}\right), \qquad \vec{\theta}_R = \left(\frac{1}{2},0,\frac{1}{2}\right)\,, \\
        s_1^2,s_2^2:  \qquad \vec{\theta}_L = \left(\frac{1}{2},0,\frac{1}{2}\right), \qquad \vec{\theta}_R = \left(0,\frac{1}{2},\frac{1}{2}\right)\,.
    \end{aligned}
\end{equation}
We can directly compute the partition function:
\begin{equation}
    \mathcal{Z}^F_{L/R}[r^2,s_1^2] = \mathcal{Z}^F_{L/R}[r^2,s_2^2] \propto 4 \vartheta_1 \vartheta_2 \vartheta_3 \vartheta_4 = 0\,.
\end{equation}
This is a non-trivial cancellation, similar to the original models \cite{Kachru:1998hd}. Nevertheless, it shows the full partition function vanishes.

For the example in \hyperref[sec:S3example]{Section 5.3} by using the presentation as $S_3 \times \Z_3$ it is clear that the commuting pairs are those involving the $\Z_3$ generator $c$, with the notation of \hyperref[fn:S3xZ3]{Footnote 9}. We therefore need to check the sectors given by a $c$ insertion in one direction, and any cyclic subgroup of $S_3$ on the other. This amounts to four different sectors: $\mathcal{Z}^F[c,\sigma_{12}]$, $\mathcal{Z}^F[c,\sigma_{13}]$,$\mathcal{Z}^F[c,\sigma_{23}]$, $\mathcal{Z}^F[c,\sigma_{123}]$. They all vanish in a subtler way, as follows. By direct computation one sees that $c$ acts trivially on $L$ movers, while $\sigma_{123}$ acts trivially on $R$ movers. What we can compute is then
\begin{equation}
   \begin{aligned}
            \mathcal{Z}^F_L[c,\sigma_{12}] &= \mathcal{Z}^F_L[1,\sigma_{12}]  \,, \quad  &\mathcal{Z}^F_L[c,\sigma_{13}] &= \mathcal{Z}^F_L[1,\sigma_{13}] \,, \\
             \mathcal{Z}^F_L[c,\sigma_{23}] &= \mathcal{Z}^F_L[1,\sigma_{23}]\, , \quad  &\mathcal{Z}^F_R[\sigma_{123},c] &= \mathcal{Z}^F_R[1,c],
    \end{aligned}
\end{equation}
which are all untwisted sector partition functions. We know that these will vanish if and only if there is a trivial eigenvalue in the fermionic action. One can check explicitly from the generators that the order 2 permutations $\sigma_{12},\sigma_{13},\sigma_{23}$ all have an eigenvalue 1 on the left, and $c$ has an eigenvalue 1 on the right.

For the example in \hyperref[sec:d6]{Section 5.4}, the only non-trivial commuting pairs in the dihedral group $D_6$ are given by $s^3$ and one of the reflections $rs^n$, in the notation of \hyperref[fn:D_6_presentation]{Footnote 10}. The vanishing of these sectors is even more subtle. The key observation is that on left movers $s^3$ acts as $(-1)^{F_L}$, by flipping the sign of all spacetime fermions. We will then compute 
\begin{equation}
    \mathcal{Z}^F_L[s^3,rs^n] = \mathcal{Z}^F_L[(-1)^{F_L},rs^n] \,.
\end{equation}
In \hyperref[app:TorusZ]{Appendix B} we show explicitly that the condition for the vanishing of this sector is that $rs^n$ has at least an eigenvalue $-1$. This can be directly checked to be true in our case for all values of $n$.

\bibliographystyle{JHEP}
\bibliography{orbifolds}

\end{document}